\documentclass[preprint]{aastex}
\usepackage{subfig}
\usepackage[dvips]{color}

\slugcomment{Accepted by ApJ}
\shorttitle{Filament Fragmentation}
\shortauthors{}

\begin{document}

\title{The Origin of OB Clusters: From 10 pc to 0.1 pc}

\author{Hauyu Baobab Liu\altaffilmark{1,2,3}}
\affil{Harvard-Smithsonian Center for Astrophysics, 60 Garden Street, Cambridge, MA 02138}
\email{hlu@cfa.havard.edu}

\author{Guillermo Quintana-Lacaci\altaffilmark{4}}
\affil{Instituto de Radioastronom\'ia Milim\'etrica, Av. Divina Pastora 7, Nucleo Central, 18012, Granada, Spain}
\email{quintana@iram.es}

\author{Ke Wang\altaffilmark{2, 5}}
\affil{Department of Astronomy, School of Physics, Peking University, Beijing 100871, China}
\affil{Harvard-Smithsonian Center for Astrophysics, 60 Garden Street, Cambridge, MA 02138}
\email{kwang@cfa.harvard.edu}

\author{Paul T. P. Ho\altaffilmark{1,2}}
\affil{Academia Sinica Institute of Astronomy and Astrophysics, \\P.O. Box 23-141, Taipei, 106 Taiwan}
\email{pho@asiaa.sinica.edu.tw}

\author{Zhi--Yun Li \altaffilmark{6}}
\affil{Department of Astronomy, \\P.O. Box 400325, Charlottesville, VA 22904}
\email{zl4h@virginia.edu}

\author{Qizhou Zhang\altaffilmark{2}}
\affil{Harvard-Smithsonian Center for Astrophysics, 60 Garden Street, Cambridge, MA 02138}
\email{qzhang@cfa.harvard.edu}

\author{Zhiyu, Zhang\altaffilmark{7}}
\affil{MPIfR/PMO}
\email{zzhang@mpifr.de}

\altaffiltext{1}{Academia Sinica Institute of Astronomy and Astrophysics}
\altaffiltext{2}{Harvard-Smithsonian Center for Astrophysics}
\altaffiltext{3}{Department of Physics, National Taiwan University}
\altaffiltext{4}{IRAM 30m}
\altaffiltext{5}{Department of Astronomy, Peking University}
\altaffiltext{6}{Department of Astronomy, University of Virginia}
\altaffiltext{7}{MPIfR}

\begin{abstract}
We observe the 1.2 mm continuum emission around the OB cluster forming region G10.6-0.4, using the IRAM 30m telescope MAMBO-2 bolometer array and the Submillimeter array.
Comparison of the Spitzer 24 $\mu$m and 8 $\mu$m images with our 1.2 mm continuum maps reveals the ionization front of an H\textsc{ii} region, the photon--dominated layer, and several 5 pc scale filaments following the outer edge of the photon--dominated layer. 
The filaments, which are resolved in the MAMBO-2 observations, show regularly spaced parsec--scale molecular clumps, embedded with a cluster of submillimeter molecular cores as shown in the SMA 0.87 mm observations. 
Toward the center of the G10.6-0.4 region, the combined SMA+IRAM 30m continuum image reveals several, parsec--scale protrusions.
They may continue down to within 0.1 pc of the geometric center of a dense 3 pc size structure, where a 200 M$_{\odot}$ OB cluster resides. 
The observed filaments may facilitate mass accretion onto the central cluster--forming region in the presence of strong radiative and mechanical stellar feedbacks. 
Their filamentary geometry may also facilitate fragmentation.  
We did not detect any significant polarized emission at 0.87 mm in the inner 1 pc region with the SMA.
\end{abstract}


\clearpage
\section{Introduction }
\label{chap_introduction}
Accretion and clustering are the most fundamental aspects of OB star formation (see Garay \& Lizano 1999, Mckee \& Ostriker 2007 and  Zinnecker \& Yorke 2007 for reviews).
OB stars are observed to form in very special regions of molecular clouds --- massive molecular clumps, typically of parsec scale in size. 
Large mass and high density are needed for the accretion flow to feed the forming massive stars as well as the associated stellar cluster at a high enough rate. 
However, how the massive clumps form, whether the internal conditions of the massive clumps facilitate the accretion process, and how the massive clumps fragment into clusters, are still open questions. 
In the present work, we improve our understanding of these issues observationally, using the Submillimeter Array (SMA; Ho, Moran, \& Lo 2004)\footnote{The Submillimeter Array is a joint project between the Smithsonian Astrophysical Observatory and the Academia Sinica Institute of Astronomy and Astrophysics, and is funded by the Smithsonian Institution and the Academia Sinica.}  and the IRAM 30m telescope\footnote{IRAM is supported by INSU/CNRS (France), MPG (Germany) and IGN (Spain).}.
We focus on the well studied UC H\textsc{ii} region G10.6-0.4, for which a wealth of complementary data are available to help address these questions. 

G10.6-0.4 is a well studied OB cluster forming region at a 6 kpc distance (Caswell et al. 1975; Downes et al. 1980). 
In the central $<$1 pc region, high bolometric luminosity (9.2$\times$10$^{5}$ L$_{\odot}$) and bright free-free continuum emission ($\ge$2.6 Jy within a 0.05 pc radius, at 1.3 cm band) were detected (Ho \& Haschick 1981; Sollins et al. 2005; Sollins \& Ho 2005), suggesting that a cluster of O-type stars has formed (O6.5--B0; Ho and Haschick 1981).
Observations of the 20 cm continuum emission using the NRAO\footnote{The National Radio Astronomy Observatory is a facility of the National Science Foundation operated under cooperative agreement by Associated Universities, Inc.} Very Large Array (VLA) unveil a $\sim$5 pc scale ionized bubble to the north of the brightest OB cluster, which suggests that the evolution of the cloud can be affected by the impact of the H\textsc{ii} region projected close to G10.6--0.4 (Ho, Klein \& Haschick 1986). 
Interferometric observations with high resolutions detected groups of water and OH masers (Genzel \& Downes 1977; Ho \& Haschick 1981; Ho et al. 1983; Hofner \& Churchwell 1996; Fish et al. 2005), and multiple high velocity $^{12}$CO outflows (Liu, Ho \& Zhang 2010), which indicates on--going and in situ star formation. 
Observations of various molecular transitions  (Ho \& Haschick 1986; Keto, Ho, \& Haschick 1987; Keto, Ho, \& Haschick 1988; Guilloteau et al. 1988; Omodaka et al. 1992; Ho, Terebey, \& Turner 1994; Klaassen et al. (2009); Liu et al. 2010; Liu, Zhang \& Ho 2011; Beltr{\'a}n et al. 2011) suggest that the general motion of the molecular gas at the 1 pc scale is dominated by gravity and shows rotation along a flattened geometry.  
The overall geometry and the excitations of the molecular gas in the central $\sim$1 pc region resemble a scaled up low--mass star forming core (Liu, Zhang, \& Ho 2011), implying that the global contraction is efficient.   
A similar geometry in the central 1 pc scale is also resolved in the luminous (L$\sim$6.6$\cdot$10$^{5}$ L$_{\odot}$) massive cluster forming region G20.08-0.14 N (Galv\'an-Madrid et al. 2009).

The high resolution spectral line observations further show the clumpiness and the spatial asymmetry of the mass and the velocity field at a scale of 0.5 pc (Liu et al. 2010; Liu, Zhang \& Ho 2011). 
This asymmetric flow appears to have a spin--up rotational motion, and might continue to the highly clumpy 0.1 pc scale, flattened, hot toroid, which was first resolved by the NH$_{3}$ optical depth studies.
The hot toroid has a high temperature and should immediately surround the central 200 M$_{\odot}$ OB cluster (Sollins \& Ho 2005; Liu et al. 2010;  Beltr{\'a}n et al. 2011), suggesting that massive stars may accrete from the very clumpy and geometrically thick molecular materials.
Observations of the radio recombination lines suggest that part of the material that is ionized by the embedded OB stars can continue across the H\textsc{ii} boundary, to feed the embedded OB stars (Keto 2002; Keto \& Wood 2006). 

Despite the rich literatures in the studies of the kinematics in the inner 1 pc region, how the parsec scale massive molecular clump connects to the more extended structure, how the massive clumps form, and how the detailed morphology and dynamics lead to the fragmentation and the formation of a cluster, remain unanswered questions.  
To improve the understanding in these aspects, we observed the 1.2 mm continuum emission in a 10 pc scale area using the IRAM 30m telescope. 
At such a large scale, the 1.2 mm continuum emission is dominated by the thermal dust emission. 
The large dish of the IRAM 30m telescope allows us to resolve the dust emission, and simultaneously provides the short--spacing information, to complement the SMA data. 
This yields high resolution and high dynamic range images for our morphological studies. 
In addition, we observed the dust polarization properties at the 0.87 mm wavelength using the SMA, to gauge the role of the magnetic field in the formation of the massive molecular clump. 
For a few 1.2 mm clumps resolved in the IRAM 30m observations, we followed up with the SMA at 0.87 mm, in order to resolve their internal structures and subsequent fragmentation.

We also compare the large scale 1.2 mm continuum emission with the Spitzer MIPSGAL 24 $\mu$m map and the GLIMPSE 8 $\mu$m map.
The diffuse 24 $\mu$m and 8 $\mu$m emissions mainly trace the ionized gas and the photon--dominant regions (PDR) at the boundary of the neutral material. 
These comparisons help to define the locations and the effects of both the embedded and the external ionizing sources. 
We retrieved the archived VLA 20 cm continuum data and the 3.6 cm and 1.3 cm continuum data to trace the diffuse and the confined ionized gas in the observed area. 

The observations and data reductions are introduced in  Section \ref{chap_obs}.
The observational results are presented in Section \ref{chap_result}.
The physical implications of the results are discussed in Section \ref{chap_discussion}. 
A brief summary and our plan of the follow up researches are given in Section \ref{chap_summary} and Section \ref{chap_future}.
We will follow up the studies of the kinematics in a future publication.
We use the coordinate system of Epoch J2000 throughout the present paper.

\section{Observations and Data Reduction} 
\label{chap_obs}
\subsection{The Millimeter and Submillimeter Continuum Data}
\paragraph{IRAM 30m Telescope}
We perform a 7$'$$\times$7$'$ OTF scanning observation using the IRAM 30m telescope MAMBO-2 array receiver, on 2011 January 20. 
The observation is carried out at a frequency of 250 GHz (1.2 mm), with a primary beam size of 10.5$''$.
We made three maps (map--1, 2, 3 hereafter), with one hour on--source time for each, centered at R.A.=18$^{h}$10$^{m}$30$^{s}$.14, Decl.=-19$^{o}$55$'$29$''$.70, using the wobbler switching mode.
The wobbler throw is 120$''$; the scanning directions of these three maps have different position angles relative to the R.A. axis, due to the rotation of the sky.
Basic data reduction was carried out using the \texttt{Mopsic} software.
The map--1 is observed in a low sky noise condition, while map--2 and map--3 showed higher noise and stripes in the image before the sky noise subtraction was performed. 

Without performing the sky noise subtractions\footnote{The sky noise subtraction algorithm subtracts the correlated (sky) emission, which significantly improves the image quality for most of the cases. In our case, it significantly underestimates the extended flux by $\sim$50 \%. We use the sky noise subtracted maps as a reference for the distribution and the geometry of the faint structures. However, we do not present the sky noise subtracted map to avoid the confusions in the brightness distributions.}, we consistently obtain the peak flux of 9.5 Jy\,beam$^{-1}$ for all three maps. 
This value is comparable with the previous SMA measurement of $\sim$10 Jy at 1.3 mm, in the central 10$''$ region (Liu et al. 2010).

A few percents of errors in estimating the flux distribution in the brightest central region will couple to the uncertainties in the modeling of the local zero flux level.
This can propagate significant ($\sim$100 mJy\,beam$^{-1}$) repetitive artifacts in the scanning direction.
We formed the final image based on map--1. 
We used the measurements in map--2 and map--3 to amend the pixels in map--1 which are corrupted by the repetitive artifacts. 
However, because of the sky subtraction effects noted above, the replaced pixels may still bias the local brightness distribution in certain areas. 
These effects, however, are limited to a level that does not affect the discussions in the present paper. 
The rms noise level of map--1 (without sky noise subtraction) is about 7 mJy\,beam$^{-1}$. 

\paragraph{SMA}
We performed the 1.3 mm band observations using the 6 meter dishes of the SMA in the subcompact configuration, the compact configuration and the very extended configuration on 2009 February 09, 2009 June 10 and 2009 July 12, respectively.
In the compact configuration and the very extended configuration observations, the observing frequencies were centered on 231 GHz (1.30 mm) in the lower sideband, and centered on 241 GHz (1.24 mm) in the upper sideband, respectively; in the subcompact configuration observation, the observing frequencies of the two sidebands were centered on 221 GHz (1.36 mm) and 231 GHz (1.30 mm).
These observations were carried out with 8 antennas. 
The pointing center of these observations is R.A.=18$^{h}$10$^{m}$28$^{s}$.683, Decl.=-19$^{o}$55$'$49$''$.07.
The primary beam size of these observations is 55$''$.
The basic calibrations were carried out using the \texttt{MIR IDL} and the \texttt{Miriad} software package; the self-calibrations of these data were carried out using the \texttt{AIPS} package.
We constructed the continuum band visibility data at 1.3 mm from the line--free channels. 
The combined SMA data sets cover a uv--sampling range of 5--410 k$\lambda$.
We detect $\sim$14 Jy of Stokes--I emission in the field of view of the SMA observations. 
The continuum image of this combined SMA data set ($\theta_{maj}$$\times$$\theta_{min}$ = 0$''$.79$\times$0$''$58) has been published in Liu, Ho \& Zhang (2010).

We performed the follow--up 0.87 mm band observations using the SMA on 2011 March 15 in the subcompact array configuration with 7 antennas, and on 2011 March 25 in the compact array configuration with 8 antennas.
The observing frequencies of the two sidebands were centered on 336 GHz (0.89 mm) and 348 GHz (0.86 mm).
The pointing center of these observations are R.A.=18$^{h}$10$^{m}$41$^{s}$.10, Decl.=-19$^{o}$57$'$41$''$.30 (P1 Region hereafter), and R.A.=18$^{h}$10$^{m}$36$^{s}$.80, Decl.=-19$^{o}$57$'$03$''$.20 (P2 Region hereafter).
The observations were carried out in the snapshot mode, with 20 minutes on--source integration per pointing in the subcompact configuration, and 16 minutes on--source integration per pointing in the compact configuration (after flagging the bad data).
The theoretical rms noise level after combining the subcompact array and the compact array data is about 3.4 mJy\,beam$^{-1}$.
We provide 2 versions of the continuum map.
In one version, we limit the uv--sampling range to 8--45 k$\lambda$, to optimize the sensitivity and the shape of the synthesized beam (4$''$.8$\times$3$''$.7).
Without limiting the uv--sampling range (8-80 k$\lambda$), we obtain a higher resolution 0.87 mm  continuum image with 2$''$.9$\times$1$''$.9 synthesized beam, to resolve the detailed structures in the brighter P1 region.

\paragraph{IRAM 30m Telescope + SMA} 
We approximate the SMA observations at 1.2 mm using the SMA data taken at 1.24 mm and at 1.30 mm.
The IRAM 30m observations detect $\sim$30 Jy of Stokes--I emission within the SMA primary beam. 
We convert the IRAM 30m image into a uv--visibility data set by using the \texttt{Miriad} tasks \texttt{demos}, \texttt{uvrandom} and \texttt{uvmodel}. 
We limit the uv--sampling range of the IRAM 30m visibility to be 0--4 k$\lambda$, to fill in the missing short spacing information in the SMA data.
We note that in such a small uv--sampling range, the visibility model of the IRAM data is minimally affected by the attenuation of the single dish primary beam. 
We have inspected the data and confirm the consistent absolute flux levels of the SMA data and the visibility model of the IRAM 30m data. 
We limit the uv--sampling range of the SMA data to be within 0--220 k$\lambda$ to optimize the sensitivity to the extended structures and the angular resolution.
By applying \texttt{invert} and (non--box) \texttt{clean} to the SMA and the IRAM 30m visibility data sets with 5000 iterations in \texttt{Miriad}, we obtain a synthesized beam of 3$''$.4$\times$3$''$.1 (P.A.=15$^{o}$), and an rms noise level of 5 mJy\,beam$^{-1}$.
The recovered flux by \texttt{clean} is 29.93 Jy. 
The final combined 1.2 mm image is formed by the \texttt{Miriad} task \texttt{restor}.

\subsection{The 0.87 mm polarization observation}
\label{chap_0p87}
We performed observations in the 0.87 mm band in polarization mode using the SMA, in the compact--north array on 2010 July 31, and in the extended array on 2010 September 12.
The observing frequencies of the two sidebands were centered on 336 GHz (0.89 mm) and 348 GHz (0.86 mm).
The pointing center of both observations is R.A.=18$^{h}$10$^{m}$28$^{s}$.683, Decl.=-19$^{o}$55$'$49$''$.07.
The primary beam size of these observations is 40$''$.
These observations cover the uv--sampling ranges of 10--135 k$\lambda$ and 20--220 k$\lambda$, respectively.
The recovered Stokes--I fluxes are 15 Jy and 10 Jy, respectively.  
Calibrations and imaging were carried out using the \texttt{Miriad} software package. 

We averaged the line--free channels in the individual observations to generate the continuum channels, and jointly imaged the upper 4 GHz and the lower 4 GHz sideband.
For the compact--north array observations, we obtain an rms noise level of 2.7 mJy\,beam$^{-1}$, for the Stokes--Q and the Stokes--U images.
For the extended array observations, we obtain an rms noise level of 3.0 mJy\,beam$^{-1}$, for the Stokes--Q and the Stokes--U images.
Jointly imaging the data from these observations yield a synthesized beam of 1$''$.2$\times$1$''$.0, and the rms noise level of 2.3 mJy\,beam$^{-1}$ for the Stokes--Q and the Stokes--U images. 

\subsection{The Centimeter Continuum Observations}
We retrieved the archived 3.6 cm continuum emission toward G10.6-0.4 in the VLA A--configuration including the VLBA Pie--Town antenna on 2005 January 2; and we observed the 3.6 cm continuum emission in the NRAO (Expanded) Very Large Array (VLA/EVLA)  C--configuration on 2009 July 27.
The pointing center of these observations is R.A.=18$^{h}$10$^{m}$28$^{s}$.683, Decl.=-19$^{o}$55$'$49$''$.07.
The primary beam size of these observations is 330$''$.
The basic calibrations, self-calibration, and imaging of these data were carried out using the \texttt{AIPS} package.
We combined the A--array+Pie--Town visibility data with the C-array visibility data, and applied a Gaussian taper in the uv--domain with a FWHM of 170 k$\lambda$ , yielding a 2$''$.3$\times$1$''$.5 synthesized beam with a position angle of -10.9$^{o}$.
The observed rms noise of the 3.6 cm continuum image is about 1 mJy\,beam$^{-1}$ ($\sim$5 K in terms of brightness temperature).
We note that the 3.6 cm continuum emission traces free--free emission based on spectral index measurements. 

We retrieved the archived VLA 20 cm continuum data, taken on 1984 June 13.
The primary beam size of this observation is 1833$''$ ($\sim$30.5$'$).
The basic calibrations, self-calibration, and imaging of these data were carried out using the \texttt{AIPS} package, yielding an rms noise level of 1 mJy\,beam$^{-1}$, with a 21$''$$\times$14$''$ synthesized beam (see also Ho, Klein \& Haschick 1986).

We retrieved the archived VLA 1.3 cm continuum data, taken on 2002 February 1.
The primary beam size of this observation is 2$'$.
The basic calibrations, self-calibration, and imaging of these data were carried out using the \texttt{AIPS} package, yielding an rms noise level of 0.13 mJy\,beam$^{-1}$, with a 0$''$.11$\times$0$''$.07 synthesized beam (see also Sollins et al. 2005; Sollins \& Ho 2005).

\subsection{The CS (1--0) Data}
We observed the CS (1-0) transition using the NRAO (Expanded) Very Large Array (VLA/EVLA) in the DnC--configuration, on 2009 September 27.
These observations had a continuous LST duration of 9 hours, with 20 available antennas after flagging. 
The pointing center is R.A.=18$^{h}$10$^{m}$28$^{s}$.683, Decl.=-19$^{o}$55$'$49$''$.07.
The primary beam size of these observations is 55$''$.
The observations cover the uv--sampling range of 4--245 k$\lambda$.
Continuum emission are averaged from the line-free channels and then subtracted from the line data.
Calibrations, self--calibrations, and imaging were carried out using the \texttt{AIPS} package, yielding a synthesized beam of 1$''$.5$\times$1$''$.1, and an rms noise level of 22 mJy\,beam$^{-1}$.
This CS dataset was previously published in Liu, Zhang \& Ho (2011).

\section{Results}
The observations discussed in this paper cover a tremendous amount of  molecular structures, which have a broad range of linear size scales, and are embedded in different ambient environments (e.g. different molecular volume density, geometry, morphology, gravitational potential, etc). To clarify the terminology (e.g. filaments, clumps, cores, protrusive features/structures, envelope, etc), we present a schematic model in Figure 1, which also forecasts parts of the observational results. 
This model presents a filamentary 10 pc scale, filamentary molecular cloud, which can be embedded in a giant molecular cloud. 
The local concentrations of mass in the filaments fragment into smaller and denser substructures on a shorter dynamical timescale than the timescale of the global contraction. 
Depending on the physical environments in the parent structures, the processes of fragmentation may also be hierarchical.  
We note that the scale bars in the schematic model reflect the physical size scales of the structures that we will be referring to. Part of the schematic model, the massive molecular envelope region, has been introduced in a previous publication (Liu, Zhang \& Ho 2011).

\label{chap_result}
\subsection{The Global Environment}
Figure \ref{fig_rgb} shows the 1.2 mm continuum map from the IRAM 30m observations.
In this figure, we label the significant 24 $\mu$m emission regions (a--o) as the suspected OB star forming regions.  
Most of these OB star forming regions already show 20 cm or 6 cm free--free continuum emissions (Ho, Klein \& Haschick 1986). 
The resolved UC H\textsc{ii} regions in our 3.6 cm observations are also labeled (A--E; see also Figure \ref{fig_2panel}).
The UC H\textsc{ii} region A located at the geometric center, which contains the OB cluster with 200 M$_{\odot}$, is resolved in the highest resolution VLA observations at 1.3 cm (Sollins et al. 2005; Sollins \& Ho 2005; see also Section \ref{chap_context}).
From Figure \ref{fig_rgb} we see at least 5 extended ($\ge$ 3 pc in the projected scale) massive filaments in the 1.2 mm continuum map.
Some of the extended structures may still be blended, and remain to be resolved with higher angular resolution. 

The most significant feature in Figure \ref{fig_rgb} is the clear separation between the H\textsc{ii} region in the northeast, and the neutral material in the southwest. 
A $>$10 pc scale ionized arc traced by the diffuse 24 $\mu$m emission is immediately followed by a slightly more extended 8 $\mu$m bright PDR shell southwest of it. 
Most of the dense neutral material are distributed further southwest of the PDR shell, and is traced by the 1.2 mm continuum emission. 
The neutral material is filamentary, and is extremely clumpy, with massive stars forming in localized dense structures. 
Multiple 24 $\mu$m bright H\textsc{ii} regions with significant PDR shells are seen over the $>$10 pc scale region (e.g. massive cluster a, c, d, e, f, j, m, n).
UV photons from the massive cluster o appear to photoionize a filament in the south, and generate a bow--shock shaped H\textsc{ii} region (see also Section \ref{chap_bow}).

The extended filament in the northwest seems to continue to the massive clusters e, f, and g.
How this filament is dynamically associated with the densest structure in the center of this map is uncertain. 
The rest of the extended filaments appear to connect to the geometrical center, an extremely bright 1.2 mm source with a projected scale of $\sim$2--3 pc.
A high concentration of 8 $\mu$m point sources is resolved in this central region, indicating the active fragmentation and the formation of a stellar cluster. 
This region has high extinction, and there might be more embedded massive stars which are not visible in the optical and near infrared observations.
Clusters of water masers (Hofner \& Churchwell 1996) and the high velocity $^{12}$CO (2--1) outflows (Liu, Ho \& Zhang 2010) were also reported in this region.
We observe the 1.3 mm continuum emission using the SMA, and observe the 3.6 cm free--free continuum emission and the CS (1--0) transitions using the VLA/EVLA in the center of this dense structure (Section \ref{chap_context}).

\subsection{The Geometrical Context of the Accretion Flow from the 5 pc Radius to the Central 0.05 pc}
\label{chap_context}
Figure \ref{fig_contour} shows the IRAM 30m 1.2 mm continuum map with much more detailed contours, overlaid with the combined IRAM 30m+SMA 1.2  mm continuum map in the central region. 
Assuming $\beta$=2, the gas--to--dust ratio of 100 (Lis et al. 1998), an average temperature of 30--50K, and subtracting the free--free continuum flux of 4 Jy (Liu et al. 2010), the detected 1.2 mm flux within the central 55$''$ SMA primary beam corresponds to 1.8--4$\cdot$10$^{4}$ M$_{\odot}$ of molecular mass.
We note that from the CS observations, Omodaka et al. (1992) estimated the mass to be 4$\cdot$10$^{4}$ M$_{\odot}$ in the central 30$''$ region.
We show the combined IRAM 30m+SMA 1.2 mm continuum map, and the velocity integrated CS (1--0) maps in Figure \ref{fig_2panel}, to demonstrate the structures in the inner region with higher angular resolutions.

While Figure \ref{fig_contour} shows multiple filaments on the scale of a few parsecs, in Figure \ref{fig_2panel}, we see multiple protrusions in the inner parsec scale region.
These protrusions seem to have no preferred alignment, exhibiting instead a radial configuration connecting to a common center. 
We visually identify the protrusions in the combined 1.2 mm map (Figure \ref{fig_2panel}). 
The identified structures are labeled by FL--NE, E, SE, S, SW2, SW1, NW2, NW1, respectively. 
The elongated structure FL--NW1, FL--E and FL--SW1 are consistent with the more extended molecular filaments, which were previously detected in the NH$_{3}$ (Ho \& Haschick 1986) and the CS (1--0) observations (Omodaka et al. 1992).
We note that the velocity gradient of  FL--E was marginally resolved by Ho \& Haschick (1986), and was interpreted as the envelope rotational motion.
The FL--S might have been detected in the previous C$^{18}$O observation (Ho, Terebey, \& Turner 1994).
However, we emphasize that the 1.2 mm continuum emission presented in this paper traces the thermal dust emission, without being biased by the excitation conditions and the molecular abundances. 

Some of the protrusions continue to a large scale, which can only be followed in the IRAM 30m 1.2 mm continuum map, for example, FL--NE, FL--NW1, FL--NW2.
There might be some foreshortened structures blended around the middle of the map (Figure \ref{fig_contour}), which are only marginally resolved in the IRAM 30m continuum map. 
There is also an apparent filament or protrusion, which continues from the central parsec region to a large scale, and is associated with the massive cluster c. 
We suspect another filamentary structure that continues to the large scale, which is related to the formation of the massive cluster m and n.
These suggestions can be tested with interferometric mosaic observations.  

Comparing the combined 1.2 mm continuum map with the velocity integrated CS (1--0) map (Figure \ref{fig_2panel}) suggests that some protrusions might continue directly into the central 0.05 pc, where the fast spinning hot toroid (Sollins \& Ho 2005; Liu et al. 2010; Beltr\'{a}n 2011) is immediately around the most massive ($\sim$200 M$_{\odot}$) OB cluster.
We note the filamentary and clumpy nature of the massive molecular envelope traced by the CS (1--0) map has been highlighted in Liu, Zhang \& Ho (2011).
As an example of the continuation of the filaments, we zoom in on the FL--SE region and show the combined 1.2 mm continuum map and CS (1--0) map in Figure \ref{fig_filament}. 
In the figure, we see that both the 1.2 mm emission and CS (1--0) emission show a number of elongated structures.  
The most prominent one follow the suggested plane--of--rotation (Keto, Ho \& Haschick 1988; Sollins \& Ho 2005; Liu et al. 2010).
Figure \ref{fig_freefree} demonstrates the relation between these elongated CS emission and the 1.3 cm free--free continuum emission from the centrally embedded UC H\textsc{ii} region.
We note that the arc--shaped features in the east of the 1.3 cm continuum map are explained as signatures of the externally ionized molecular clumps, which are approaching the embedded OB cluster (Sollins \& Ho 2005). 

This is the first time we resolve the geometry of the massive accretion flows in the distant (few kpc) OB cluster forming regions, on such a large scale and with such a high spatial dynamic range. 
However, the resolved structures mentioned here are all projected onto the two dimensional images. 
High resolution spectral line observations (Liu et al. 2010; Liu, Zhang \& Ho 2011) can be used to study the three dimensional geometrical context and the overall dynamics. 
This will be the subject in our next paper.

\subsection{The Millimeter Clumps and the Submillimeter Cores}
\label{chap_mm}
The parsec scale filaments have complicated internal structures. 
As an example, Figure \ref{fig_flrg} shows the IRAM 30m 1.2 mm continuum map in FL--E, overlaid with the saturated 24 $\mu$m and 8 $\mu$m \texttt{Spitzer} images. 
From this figure, we see that both the 1.2 mm continuum emission and the near infrared opacity suggest a structure continuing from the southwest, which may still contain substructures. 
More detailed structures are suggested by the near infrared opacity, although it is hard to distinguish if those structures are physically associated or a mere projection along the line--of--sight.
In such an evolved region, the distribution of the foreground and background extended infrared emissions severely confuse the structure identification. 
The local structures can best be recognized from the millimeter and submillimeter thermal dust emission.
However, the IRAM 1.2 mm continuum map has much poorer angular resolution as compared to the 24 $\mu$m and 8 $\mu$m \texttt{Spitzer} images. 

Here, we define the millimeter clumps as compact structures identified in the IRAM 1.2 mm map, which have sizescales of $\sim$0.5 pc.
We define the submillimeter cores as compact structures identified in the SMA 0.87 mm Stokes--I image, which have sizescales $\lesssim$0.2 pc.
We identify the millimeter clumps as well as the submillimeter cores in regions where the fluxes are predominantly contributed by FL--E and FL--S, since structures in other regions, particularly in the east appear to be blended due to the still insufficient angular resolution. 
In addition, the millimeter and the submillimeter continuum emissions in FL--E and FL--S are minimally contaminated by the free--free continuum emission, and are therefore good tracers of the molecular mass (Figure \ref{fig_contour}, \ref{fig_ionized}; see also Ho, Klein \& Haschick 1986 for the free--free continuum emission). 
Note that these observationally defined features are still limited by the insufficient angular resolution and sensitivity of our observations, with likely unresolved substructures.

We perform 2--dimensional Gaussian fit using the \texttt{AIPS} task \texttt{IMFIT} to obtain the size. 
We note that the size from the fitting is limited by the (synthesized) beam of the observations.
The Gaussian fits of the fainter cores are biased by the confusion with the emission from the filaments. 
The profiles of the brightness distribution of the identified structures may be non--Gaussian due to the internal structures. 
Hence, to obtain the total flux, we simply sum the flux within the ellipses as defined by the Gaussian fits. 

We provide the preliminary mass estimates based on the optically thin dust emission formulation in Lis et al. (1998), assuming an average temperature of 25K, the gas--to--dust ratio of 100, and $\beta$=2.
We approximate the volume of each of the clumps and cores by assuming a spherical geometry, with the radius being the average of their projected major and minor axes. 
The estimated flux and mass of the clumps and cores can be biased to be factors of 2--3 higher, due to those uncertainties in the Gaussian fits.
However, while estimating the density, these biases will be compensated by the similar bias in the volume.
The coordinates, size, millimeter or submillimeter flux, mass, and the average density are summarized in Tables \ref{table_mm} and \ref{table_submm}.
With the similar assumptions, the overall molecular mass in FL--E and FL--S, without including the structures in the central 3 pc dense region, are 6$\cdot$10$^{3}$ M$_{\odot}$ and 4.5 $\cdot$10$^{3}$ M$_{\odot}$, respectively. 

The identified structures are also labeled (MM 1--9, SMM 1--6, and SMM 1--5 a, b, c) in Figures \ref{fig_contour}, \ref{fig_ionized}, and \ref{fig_condensations}.
The millimeter clumps in FL--E have an average projected separation of 0.82 pc; and the millimeter clumps in FL--S have an average projected separation of 0.54 pc.
Our follow--up submillimeter Stokes--I observations focus on the millimeter clumps MM1, MM2 and MM4.
We resolved 6 submillimeter substructures in the lower resolution versions of the 0.87 mm Stokes--I image (SMM 1--6, Figure \ref{fig_ionized}).
SMM 1 and SMM 2 have a projected separation of 0.27 pc; and SMM 3, 4, 5 have an average projected separation of 0.15 pc.
We note that the Jeans length is 0.14 pc, assuming an average temperature of 25 K, and an average density of n$_{H_{2}}$ = 10$^{5}$ cm$^{-3}$.

Without limiting the uv--sampling (Section \ref{chap_0p87}), we made a higher angular resolution (2$''$.9$\times$1$''$.9) 0.87 mm Stokes--I image.
We find that the core SMM1 is resolved into three independent submillimeter cores, while SMM 2, 3, and 5 are also each resolved into two further submillimeter cores (Figure \ref{fig_condensations}).  
This implies smaller actual sizes and average separations of submillimeter cores in MM1 and MM2.
We label the resolved substructures by SMM 1--5 with suffix a, b, c.
The submillimeter cores SMM 1a, 1b, 1c, 2a and 2b seem to follow one elongated distribution; the submillimeter cores SMM 3a, 3b, 4, 5a, 5b are aligned in another elongated distribution. 
With the still insufficient angular resolution of the SMA 0.87 mm Stokes--I image (Section \ref{chap_0p87}), we may underestimate the volume density (Table \ref{table_submm}) of the submillimeter cores. 
The submillimeter core SMM 6 has an elongated shape (Figure \ref{fig_ionized}), which may indicate further internal structures, but is too faint to be resolved.

\subsection{Dust Polarization}
\label{chap_pol}
We do not detect the Stokes--Q and Stokes--U emissions at the 3$\sigma$ level, in any combination of the compact--north array and the extended array data.
We note that our polarized intensity maps have achieved comparable resolutions and lower rms noise levels as compared with the previous SMA studies of distant OB star formation regions (G31.41+0.31: Girart 2009; G5.89-0.39: Tang et al. 2009a; W51 e2/e8 cores: Tang et al. 2009b).
Our non--detection suggests that G10.6--0.4 is less than 2.8\% polarized.
Possible explanations for the lower level of the polarized emission and the physical implications are discussed in Section \ref{chap_mag}.

\subsection{The Ionized Bubble and the Patchy Surrounding Molecular Gas}
\label{chap_patchy}
The early VLA 20 cm observations show an ionized bubble to the north of the central OB cluster. 
This ionized bubble can be seen in the 24 $\mu$m emission (Figure \ref{fig_rgb}), and also with the 8 $\mu$m bright PDR shell external to it.
The 24 $\mu$m emission additionally shows some patchy structures on the eastern edge of the ionized bubble.
Those patchy structures have high opacities, and are dark in the 24 $\mu$m and 8 $\mu$m maps.

The observed area has strong diffuse infrared emission.
The structures in the background infrared brightness distributions and the foreground infrared emission can confuse the identifications of structures. 
We report the patchy dust emission from the same area in the 1.2 mm band.
A blow--up view of the IRAM 30m 1.2 mm continuum map overlaid with the VLA 20 cm continuum map is provided in Figure \ref{fig_bubble}.

The distribution of the 20 cm continuum emission suggests that the 1.2 mm emission from the patchy molecular gas are not contaminated by the free--free emission.
This patchy molecular gas shows rich subparsec scale structures. 
With our current data, how these patchy structures form, and if they are interacting with the ionized bubble or the H\textsc{ii} region, are unclear. 
We suspect that they are naked dense cores which lost their ambient gas through some interactions; or they can be some foreground, isolated, subparsec scale molecular cloudlets. 
These have to be distinguished in future observations.

\section{Discussions}
\label{chap_discussion}
\subsection{The Impulse Driven or Regulated Structure Formation?}
From Figure \ref{fig_rgb}, we see the formation of the OB stars over the entire $\ge$10 pc projected area. 
The formation of these OB stars seems to be synchronized on a time interval, which is short in the sense that it takes $\gg$10$^{7}$ years to propagate the sound wave in this $\sim$10 pc cloud.
This propagation timescale is much longer than the collapsing timescale plus the life time of the massive stars. 
The Alfv\'en waves can propagate supersonically. 
However, they will be restricted by the detailed magnetic field line configurations and the magnetic field strength. 

Are there large scale physical phenomena, which are globally regulating or triggering the local structure formations (e.g. the formation of the massive core, or the collapsing of the massive cores, etc)?
The impulse from the H\textsc{ii} region in the northeast might be a candidate of this global mechanism, which was suggested by Ho, Klein \& Haschick (1986) with their VLA 20 cm continuum observations. 
This scenario has to be tested by the numerical hydrodynamical simulations. 

\subsection{The Ionizing Pattern as an Evidence of the Filamentary Nature}
\label{chap_bow}
With the 1.2 mm dust continuum maps, one may argue that the filamentary geometry of the neutral material could be a projection effect. 
In this section, We present the case of the ionizing pattern around the parsec scale filament in the south, and the massive cluster o (Figure \ref{fig_rgb}), to argue in favor of the elongated geometry of the parsec scale structures. 
While fragmentation in filamentary infrared dark clouds are ubiquitously seen, we argue that our result provides hints for their evolutionary context.
We can not rule out the possibility that there are edge--on sheets which in projection appear like filaments.
However, it is questionable whether a $>$5 pc scale sheet is likely at such a late evolutionary stage where clusters of smaller structures and massive stars are already present. 

The idea is that if there is an OB cluster located close to the parsec scale massive filaments (with a gradient of volume density along the radius), its ionizing photons and stellar wind will be blocked in the direction of the dense neutral material.
The UV photons emanating from the OB stars will only efficiently ionize the outer edges of the filament.  If the filament presents a small solid angle to the OB cluster, most of the UV photons, the stellar wind and the ionized gas will flow around the filament, and form a \textbf{U} shaped free--free continuum emission region.
Such a \textbf{U} shape illumination pattern is seen around the massive filament in the south (Figure \ref{fig_ionized}).
If that filament were just filamentary in projection, but is an extended smoothed sheet in reality, we do not expect to see the ionized gas penetrating to the northeast of it; instead, the free--free emission should be limited to the southwest. 
Alternatively, the \textbf{U} shape illumination pattern can also be explained by the addition of two photoionized molecular filaments.
This scenario can be examined by deeper observations of thermal dust emission with higher angular resolution. 

We note that the cloud with a filamentary morphology may be self--shielded from an OB cluster deeply embedded in the geometrical center of the cloud. 
While the parsec scale filament in the south is illuminated by the massive cluster o and shows significant 8 $\mu$m emission (Figure \ref{fig_rgb}), the filament in the east remains dark at 8 $\mu$m.
If there are smaller and denser filaments continuing to the centrally embedded OB cluster, the small cross--sections of the filamentary accretion flow is minimally affected by the radiation from the luminous OB cluster, and the pressure force of the ionized gas.
Therefore, it is conducive for the accretion of the massive stars. 

Previous observations have suggested that the major, massive filaments can be formed from the merging of the smaller ones (Jim{\'e}nez-Serra et al. 2010).
The filaments may also collide with each other, which is conducive for the formation of cores and the massive stars (Galv\'an-Madrid et al. 2010).

At 10 pc scale, the filamentary morphology has a low volume filling factor, or a high porosity even after being projected onto a 2--dimensional observing area.
The intensity ratios of molecular gas tracers (e.g. HCN/CO) can then be easily converted to volume filling factors (e.g. dense gas volume filling factor) as in extragalactic studies. 
From Figures \ref{fig_contour}, \ref{fig_ionized}, and Figure \ref{fig_condensations}, we see concentration of dense gas in the form of massive molecular clumps and molecular cores. 
The volume densities of these clumps and cores (Tables \ref{table_mm}, \ref{table_submm}) are high enough to excite the dense gas tracers CS, HCO$^{+}$, HCN, and H$_{2}$CO. 

\subsection{A Geometrically Regulated Fragmentation in Parsec Scale Filaments}
The parsec scale filaments, especially the two in the south and in the east, show rich marginally resolved 1.2 mm substructures, the millimeter clumps (Figure \ref{fig_contour}). 
The clumps in the eastern filament show the generally stronger 1.2 mm  emission than those in the southern filament, which can be due to their difference in mass, temperature, or the dust properties. 
This can be further examined by observing the NH$_{3}$ emission with multiple transitions, and by observing the dust continuum emission in multiple frequency bands. 
At least three massive clumps in the eastern filament may be forming massive stars, and show bright 8 $\mu$m point sources around the region we label cluster l (Figure \ref{fig_rgb}).

Clumps in the southern and the eastern filaments seem to be regularly separated by 0.5--1 pc, implying a scale length for the local contractions which may be related to the length and width of the filaments.  
In the clumps MM1 and MM2, our follow up SMA observations have resolved much more submillimeter cores, which have an averaged projected separation comparable to the thermal Jeans Length (Section \ref{chap_mm}, Figure \ref{fig_condensations}). 
Compared with other high resolution interferometric observations of filamentary infrared dark clouds, it is likely that hierarchical fragmentation is common.
Undoubtedly, there will be more blended substructures in our IRAM 30 m map due to the still insufficient angular resolution. 
For example, in the case study of the infrared dark cloud G28.34+0.06 ($d\sim$4.8 kpc; Wang et al. 2008; Zhang et al. 2009; Wang et al. 2011), the 1.3 mm continuum observation of SMA resolved five cores within a single local maximum in the IRAM 30m 1.2 mm continuum map. 

We note that the submillimeter cores in the molecular clumps MM 1 and MM 2 (Figure \ref{fig_condensations}) are not randomly distributed, but are aligned linearly, forming the main axis of the filament. 
That the successive fragmentation at even smaller scales, continue to be aligned along the filaments, suggests a common process. 

The process of fragmentation into clumps can be regulated by the density, mass concentration, the geometry or the morphology of the clump, the local magnetic field structure, the local feedback and the ambient pressure, the velocity field, interactions, and so on. 
Successive fragmentation takes place as the local conditions change.
Furthermore, collision or merging of filaments may also be an important process. 
We postpone the detailed statistical studies of these cores until we obtain follow up observations that achieve the physical resolution of  $\ll$0.1 pc, and achieve a much better sensitivity to recover the solar mass or smaller substructures. 

Some of the parsec scale filaments appear to converge into the geometric center of a 2--3 pc scale densest structure.
The abundance of protrusions in both the IRAM 30m 1.2 mm continuum map, and the combined IRAM 30m + SMA 1.2 mm continuum map (Figure \ref{fig_contour}) support such a scenario.
We note however that the massive stars are distributed over the entire 10 pc area. 
Hence, it appears that the star formation process may be driven by local processes, and may not be related to inflow from large scales via the filaments. 

\subsection{Morphologically Regulated Cluster Formation}
\label{chap_frag}
We provide simple arguments based on the global and the local free--fall timescale, to justify that the filamentary accretion flows with a radial alignment will allow efficient fragmentation during the quick global contraction.  
Theories on how contractions depend on morphology, geometry, and initial conditions, are described by Ledoux (1951), Ostriker (1964), Larson (1985), Vishnaic (1994), Whitworth et al. (1994), Curry (2000), Myers (2009), and Myers (2011).
Our estimates specifically focus on the materials in the central 2 pc region (i.e. inside the SMA primary beam shown in Figure \ref{fig_2panel}), in which we see a concentration of bright infrared point sources (Figure \ref{fig_rgb}). 
We do not consider the magnetic field in our estimates, which is consistent with our observations that strong and organized magnetic fields may not be present at the sampled sizescales (see also Section \ref{chap_mag}).  
With some rescaling, the same calculations can be applied to the more extended area while a lack of understanding of the large scale initial turbulence brings large uncertainties. 

The free--fall timescale is inversely proportional to the square root of the averaged density $\rho$.
The value of $\rho$ can be estimated by
\begin{equation}
\rho = \frac{M(R)} {\frac{4}{3}\pi R^{3}}, 
\label{eq_rho}
\end{equation}
while $R$ is the radius around a 'local' center of mass, and $M(R)$ is the total mass enclosed in the radius $R$.
For the case of G10.6--0.4, in the central 2 pc region, we have resolved $\sim$8 protrusions (Figure \ref{fig_2panel}). 
Assuming a filamentary geometry for those protrusions, of which the width is a fraction of a parsec, say 0.1--0.3 pc, the volume occupied by each filament can be approximately estimated as the volume of a 2 pc rod, which is $(\pi/4)\cdot$$(0.1\mbox{---}0.3)$$^{2}$$\cdot2$ $=$ 0.016--0.14 pc$^{3}$. 
The volume filling factor in the inner 2 pc region can be approximately estimated by
\begin{equation}
\frac{8\cdot(0.016\mbox{---}0.14)}{\frac{4}{3}\pi\cdot(1\mbox{ pc})^{3}} = 3.0\mbox{---}27\%.
\label{eq_filling}
\end{equation}
If we project all structures along the line--of--sight, we can also estimate the surface filling factor, which is $\ge$50\%.  
We note that the smaller beam filling factor of 30\% reported in the previous low resolution NH$_{3}$ observations (Keto, Ho, \& Haschick), suggests that most of the mass are concentrated to filaments or in the central massive core.

The estimated small volume filling factor implies potentially a much higher local volume density than the averaged volume density in that region. 
Within the region enclosed by the 1 pc radius, the free--fall contraction timescale of the local perturbations can be a factor of  1.9--5.8 times shorter than the global free--fall contraction timescale.
Depending on the support of the rotational velocities, the timescale of the large scale contraction could be a few times longer than the global free--fall contraction timescale (Ho \& Haschick 1986; Keto, Ho \& Haschick 1987, 1988; Keto 1990; Liu et al. 2010; Liu, Zhang \& Ho 2011).
Thus we suggest that although the system may have a quick global contraction (see also the \textit{clump--fed} scenario in Wang et al. 2010), the existence of small scale structures implies that efficient fragmentation and subsequent star formation may have already occurred. 
The accretion flows are clearly highly clumpy, which have been resolved in the high resolution NH$_{3}$ absorption experiment (Sollins \& Ho 2005), and the PV diagram of $^{13}$CS (5--4) (Liu et al. 2010).
The fragmentation of the accretion flows is consistent with the observed clusters of UC H\textsc{ii} regions, multiple water masers and high velocity $^{12}$CO outflow sources (Figure \ref{fig_2panel}), many dusty cores with strong 1.3 mm emission, all external to the centrally located, compact, fast spinning, hot toroid (Liu, Ho \& Zhang 2010; Liu, Zhang \& Ho 2011). 
The earliest epoch of star formation will generate turbulence, which will help regulate the subsequent star formation efficiency and the accretion of the centrally embedded massive star (G10.6--0.4: Liu, Ho \& Zhang 2010; Simulations: Li \& Nakamura 2006; Nakamura \& Li 2007; Carroll et al. 2009; Wang et al. 2010).

From the three color image (Figure \ref{fig_rgb}), we see a higher concentration of the 8 $\mu$m sources in the central parsec region than in the extended region, which may be explained by the faster fragmentation and contraction of the filamentary structures in the higher density environment. 
We note that the interferometric observations of the nearby massive cluster forming region, the Orion--KL central core, have also demonstrated the highly filamentary nature of the molecular gas, as well as the fragmentation in those filaments, in the inner $\sim$1 pc region (Wiseman \& Ho 1996; Wiseman \& Ho 1998).
Note that on such a small scale, the filaments do not need to be primordial nor a continuation of the outer structures. 
The interactions between the ambient gas with the (proto--)stellar feedback can also lead to the formation of filaments in the later evolutionary stages.
Of course, the filaments can also be destroyed by the same feedback mechanisms (Wang et al. 2010). 
The process of infall and fragmentation is highly dynamical and chaotic. 

\subsection{Spin--up Signature of the Filamentary Accretion Flow}
\label{chap_spiral}
The \textit{red excess} of the velocity field in G10.6-0.4 was previously reported in the NH$_{3}$ absorption line experiments, which is consistent with a high velocity infall (Ho \& Haschick 1986; Keto, Ho \& Haschick 1987, 1988; Keto 1990; Sollins \& Ho 2005).
From 1$''$ resolution emission line observations, Liu et al. (2010) confirms the red excess in the central $\sim$0.06 pc (2$''$) region, and further sees the high spatial asymmetry and the clumpiness in the mass distribution, from the position--velocity (PV) diagram.
Motivated by the resolved protrusions in the present work (Figure \ref{fig_2panel}), we propose that the spin--up rotational motions of the accretion flow with a filamentary morphology may coherently explain the red excess, the spatial asymmetry and the clumpiness in those PV diagrams. 
The detailed modeling will be provided in our follow--up kinematics studies.

\subsection{The Magnetic Field}
\label{chap_mag}
SMA observations of luminous massive star forming regions  (G31.41+0.31: Girart 2009; G5.89-0.39: Tang et al. 2009a; W51 e2/e8 cores: Tang et al. 2009b) have shown that the polarization percentage of the 0.87 mm thermal dust emission, range from a few percent (e.g. 4\% in G31.41+0.31), to up to $\sim$8\% in the W51 e2/e8 regions, and up to $\sim$22\% in the G5.89-0.39 region.
The polarization observations of G10.6-0.4 achieved an improved sensitivity over the previous SMA studies of similar targets.  
While those previous cases all show significant detections of polarized flux, in the present work, we do not detect the polarized emission above the 3$\sigma$ rms noise level. 
In G10.6-0.4, the polarization percentage of the dust emission at 0.87 mm is constrained to be less than 2.8\%. 
Such a polarization percentage is lower than the reported polarization percentages in those previous cases. 
If G10.6-0.4 is a strongly magnetized case, the low polarized flux can be interpreted by some differences in the dust properties from the other targets, which produce less polarized flux, the relatively inefficient dust alignment in G10.6--0.4, and/or the canceling of the polarized emission from structures overlapped along the line--of--sight.
The feedback of the high velocity molecular outflows and the expansions of the H\textsc{ii} regions may, but not necessarily disturb the magnetic field and lead to the non--detection of the polarized flux. 
In the case of G5.89-0.39, the embedded UC H\textsc{ii} region has already undergone an expansion phase, and multiple molecular outflows have been detected, and the magnetic field structure is strongly affected (Tang et al. 2009a).
The magnetic field might also be more important on a much smaller scale, which is unresolved in our SMA observations.  

The radial morphology of the filamentary structures in the central core, suggests contraction by gravity to be dominant over all possible sources of support.
This needs to be confirmed via modeling of the kinematics.
The absence of detectable magnetic field structures is consistent with this scenario. 

\section{Summary}
\label{chap_summary}
Comparing our high resolution 1.2 mm continuum maps and the CS (1--0) observations, suggests that the massive molecular clump containing a luminous ($\gg$10$^{5}$ L$_{\odot}$) OB cluster is part of a $\ge$10 pc scale organized structure.
The extended molecular gas has a filamentary geometry, starting from a $\sim$5 pc extent in radius, to a geometrical center of a 2--3 pc scale dense structure. 
The parsec scale filaments show regularly spaced molecular clumps, each containing massive molecular cores that are associated with the local massive star formations. 
Part of these parsec scale filaments might directly continue into the central 0.05 pc radius hot toroid, which encircles the highest mass stellar cluster.
In this central region, the radial filamentary configuration suggests accretion flows which would allow simultaneously a quick global contraction and the efficient local fragmentation. 
Such a model would be consistent with the previously seen centrally located 0.1 pc scale fast--spinning hot toroid, and the rich cluster of water masers, millimeter cores, and high velocity $^{12}$CO (2--1) outflows. 
We propose that the spin--up rotational motions of a filamentary accretion flow can explain the red/blue excess of the velocity field, and the spatial asymmetry of the mass distributions, although the interpretation is not unique. 

In this scenario, the magnetic field may not, and does not necessarily play an important role on the scale that we are observing (0.05--5 pc radius).
The large numerical hydrodynamics simulations to compare the strongly and the weakly magnetized cases will help to improve the physical insights, and also improve the interpretation of the data.
The proposed scenario in this paper might not be the unique mode of OB cluster formation. 
Survey observations of a larger sample will be useful to identify or to classify other modes of OB cluster formation. 
We emphasize that the subject under the consideration in the present paper is specifically the luminous ($\gg$10$^{5}$ L$_{\odot}$) OB cluster. 

\section{Future Work}
\label{chap_future}
In the present work, we suggest the convergence of the massive filamentary structures is one very important aspect in OB cluster formation. 
However, how such filaments of a few parsec scale form, is still uncertain, and should be a very fundamental aspect in the study of the interstellar medium. 
The very sensitive high resolution and high dynamic range spectral line and dust polarization observations might help to address this question.

Another aspect which cannot be addressed by our present research is how the 0.1 pc scale hot toroid converge to form the embedded OB clusters. 
High resolution molecular and recombination line observations with ALMA will certainly address this issue. 

\acknowledgments
We thank the SMA staff for making these observations possible.
We also thank the IRAM staff for help with the observations and for developing the data reduction routines.
Liu thanks ASIAA for supporting this research. 
This work is based in part on observations made with the \textit{Spitzer Space Telescope}, which is operated by the Jet Propulsion Laboratory, California Institute of Technology. 
We thank the GLIMPSE team (PI: E. Churchwell) and MIPSGAL team (PI: S. Carey) for making the IRAC and MIPS images available to the community. 
K. W.  acknowledges the support from the SMA predoctoral fellowship and the China Scholarship Council.

{\it Facilities:} \facility{SMA, VLA, EVLA, IRAM 30m Telescope, Spitzer Space Telescope}


\begin{table}[h]
\footnotesize{
\begin{tabular}{ccccccc}
Object	&	R.A. (J2000)		&		Decl. (J2000)		& 	Size (arsec$\times$arsec)		&		Flux (mJy)		& H$_{2}$ Mass (M$_{\odot}$)		&	Volume Density (10$^{5}$ n$_{H_{2}}$/cm$^{-3}$)\\\hline\hline
MM1 & 18$^{h}$10$^{m}$41$^{s}$.59 & -19$^{o}$57$'$41$''$.4	&	15$\times$13			&		168				&				263					&		1.5									\\
MM2 & 18$^{h}$10$^{m}$40$^{s}$.41 & -19$^{o}$57$'$41$''$.4	&	15$\times$13			&		154				&				241					&		1.4									\\
MM3 & 18$^{h}$10$^{m}$38$^{s}$.61 & -19$^{o}$57$'$21$''$.7	&	27$\times$19			&		335				&				525					&		0.68									\\
MM4 & 18$^{h}$10$^{m}$36$^{s}$.84 & -19$^{o}$57$'$03$''$.8	&	15$\times$12			&		122				&				191					&		1.2									\\
MM5 & 18$^{h}$10$^{m}$34$^{s}$.61 & -19$^{o}$58$'$42$''$.2	&  27$\times$16			&		242				&				379					&		0.60									\\
MM6 & 18$^{h}$10$^{m}$34$^{s}$.61 & -19$^{o}$58$'$24$''$.7	&	25$\times$16			&		224				&				351					&		0.69									\\
MM7 & 18$^{h}$10$^{m}$33$^{s}$.86 & -19$^{o}$58$'$17$''$.7	&	27$\times$21			&		354				&				555					&		0.63									\\
MM8 & 18$^{h}$10$^{m}$33$^{s}$.37 & -19$^{o}$58$'$00$''$.2	&	30$\times$19			&		351				&				550					&		0.59									\\
MM9 & 18$^{h}$10$^{m}$32$^{s}$.62 & -19$^{o}$57$'$39$''$.2	&	27$\times$14			&		192				&				301					&		0.55									\\\hline
\end{tabular}
}
\caption{The millimeter clumps identified from the IRAM 30m MAMBO-2 1.2 mm continuum observation.}
\label{table_mm}
\end{table}

\begin{table}[h]
\footnotesize{
\begin{tabular}{lcccccc}
Object	&	R.A. (J2000)		&		Decl. (J2000)		& 	Size (arsec$\times$arsec)		&		Flux (mJy)		&	H$_{2}$ Mass (M$_{\odot}$)		&  Volume Density (10$^{5}$ n$_{H_{2}}$/cm$^{-3}$)\\\hline\hline
SMM1a & 18$^{h}$10$^{m}$41$^{s}$.75 & -19$^{o}$57$'$38$''$.2	&	3.4$\times$2.0			&		56			    &				37					 	&		30	\\
SMM1b & 18$^{h}$10$^{m}$42$^{s}$.04 & -19$^{o}$57$'$39$''$.8	&	3.5$\times$2.7			&		28			    &				19					 	&		10	\\
SMM1c & 18$^{h}$10$^{m}$41$^{s}$.97 & -19$^{o}$57$'$36$''$.5	&	3.7$\times$1.7			&		22			    &				15					 	&		12	\\									
SMM2a & 18$^{h}$10$^{m}$41$^{s}$.52 & -19$^{o}$57$'$38$''$.7	&	2.9$\times$2.3			&		24				&				16					&		14									\\
SMM2b & 18$^{h}$10$^{m}$41$^{s}$.31 & -19$^{o}$57$'$38$''$.7	&	2.7$\times$1.9			&		20				&				13					&		16									\\
SMM3a & 18$^{h}$10$^{m}$40$^{s}$.76 & -19$^{o}$57$'$40$''$.9	&	3.4$\times$2.9			&		49				&				32					&		25									\\
SMM3b & 18$^{h}$10$^{m}$40$^{s}$.98 & -19$^{o}$57$'$44$''$.3	&	5.0$\times$2.7			&		45				&				30					&		8.2									\\
SMM4 & 18$^{h}$10$^{m}$40$^{s}$.53 & -19$^{o}$57$'$38$''$.9	&	3.6$\times$2.6			&		37				&				25					&		13									\\
SMM5a & 18$^{h}$10$^{m}$40$^{s}$.36 & -19$^{o}$57$'$37$''$.5	&	5.9$\times$2.7			&		69				&				46					&		9.1									\\
SMM5b & 18$^{h}$10$^{m}$40$^{s}$.22 & -19$^{o}$57$'$36$''$.1	&	3.8$\times$2.2			&		26				&				17					&		9.9									\\

SMM6 & 18$^{h}$10$^{m}$40$^{s}$.43 & -19$^{o}$57$'$36$''$.7	&	6.9$\times$5.0			&		43				&				28					&		2.1									\\\hline
\end{tabular}
}
\caption{The submillimeter cores identified from the SMA 0.87 mm Stokes--I images. We note the volume densities in the last column are the averaged values for unresolved structures, which can be underestimated.}
\label{table_submm}
\end{table}

\clearpage

\begin{figure}
\resizebox{15cm}{!}{\includegraphics{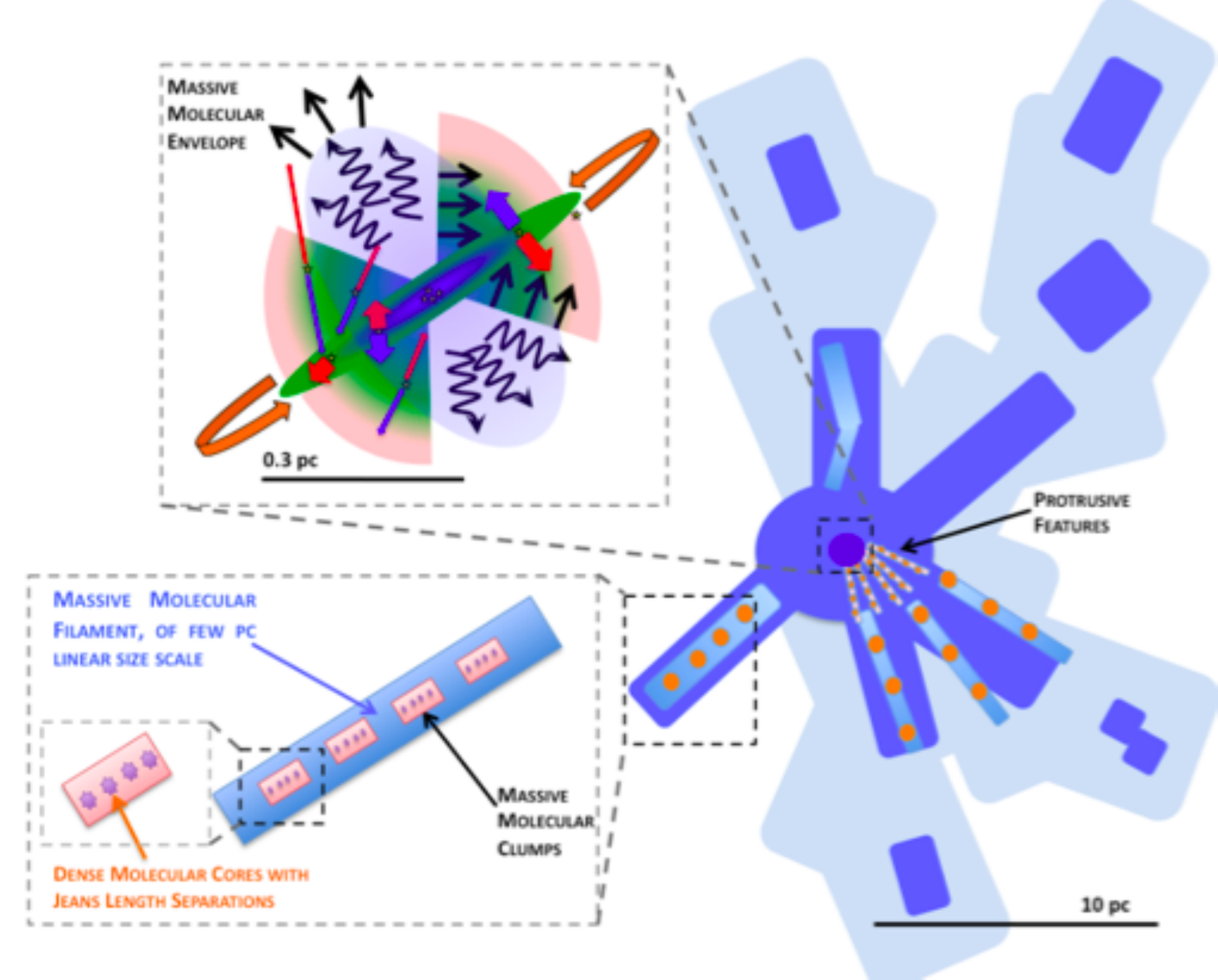}}
\caption{The schematic model of the observed molecular structures. This model presents a scenario of the hierarchical structure formation in a filamentary, OB cluster forming molecular cloud. The parent molecular cloud is shown in light--blue color. Denser, massive molecular filaments at few parsec scale, and some local dense structures are shown in blue color. Even denser molecular filaments that have embedded massive molecular clumps are shown in blue gradient. The embedded massive molecular clumps are shown by the orange circles. 
The blow--up of a filament further shows the dense molecular cores, which are the internal structures embedded in the massive molecular clumps.
In this model, the dynamical processes (e.g. fragmentation, contraction) in denser regions are occurring on shorter ($\sim$free--fall) dynamical timescales, and have smaller ($\sim$Jeans) separations and size scales. The massive molecular envelope embedded in the central region, is affected by many high velocity molecular outflows (as indicated by the blue and the red envelopes), the ionizing photons (wiggle arrows), the pressure forces of the ionized gas (black arrows), and the rotational motions (as indicated by the orange arrows). For more descriptions about the schematic model of the massive molecular envelope, and the observing data, see Liu, Zhang \& Ho (2011).
We note that the massive molecular envelope is clumpy and filamentary, and appears to be connected to the extended structures via several high density, protrusive features.}
\label{fig_schematic}
\end{figure}

\begin{figure}
\hspace{-2cm}
\resizebox{\columnwidth}{!}{\includegraphics{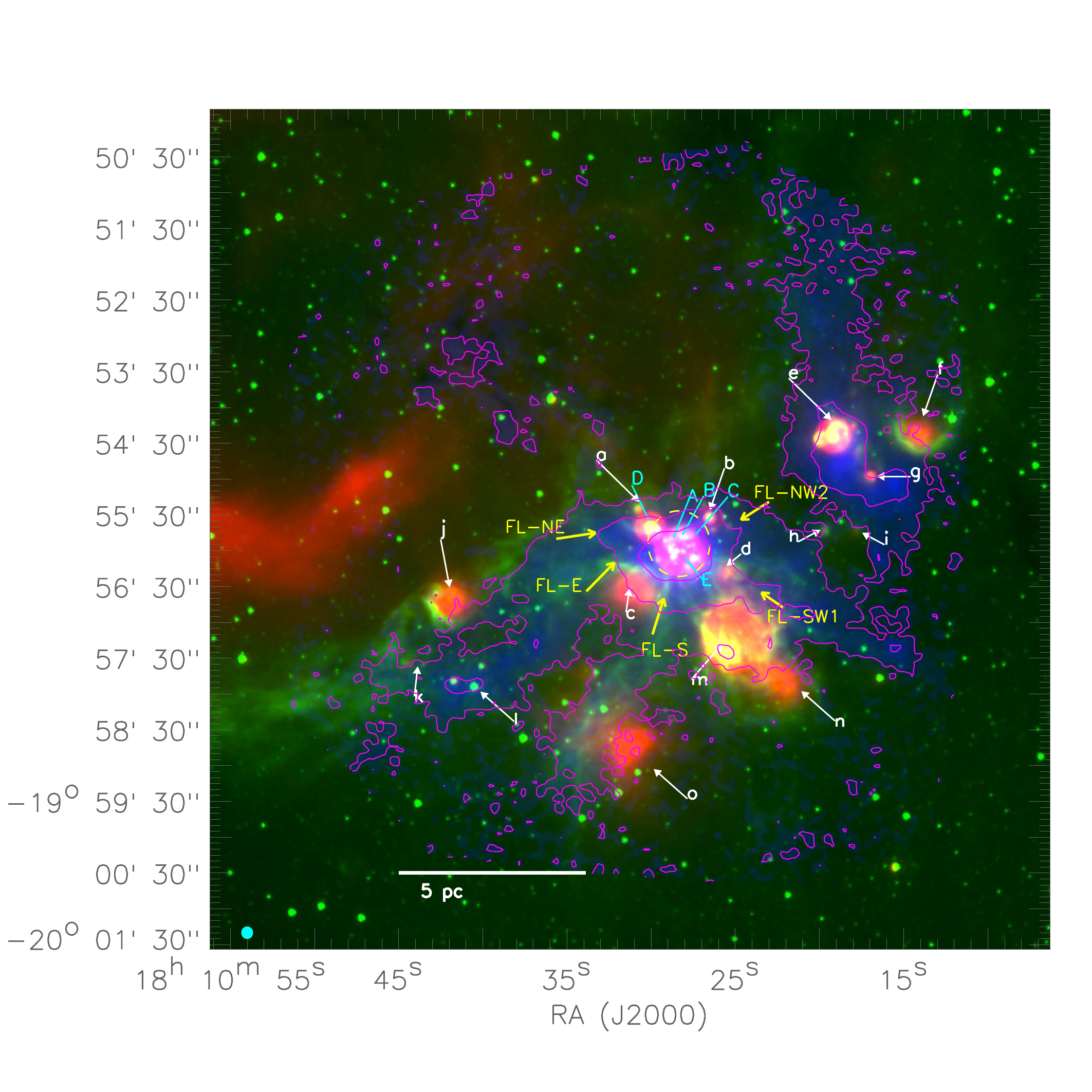}}
\vspace{-1.0cm}
\caption{The Spitzer MIPSGAL 24 $\mu$m image (red), the Spitzer GLIMPSE survey 8 $\mu$m image (green), and the IRAM 30m 1.2 mm continuum image (blue color and contours). The beam of 1.2 mm continuum image ($\theta_{maj}$$\times$$\theta_{min}$=10$''$.5$\times$10$''$.5) is shown in the bottom left corner. A scale bar indicating the 5 pc projected length is shown, with the assumption of a 6 kpc line--of--sight distance. The suspected OB clusters associated with the bright 24 $\mu$m emission are labeled (a--o) with white arrows. The dashed circles represent the primary beam of the SMA 1.3 mm observations.  The UC H\textsc{ii} regions in the SMA primary beam are labeled (A--E) with cyan solid arrows (see also Figure \ref{fig_2panel}). Five extended elongated structures or filaments are labeled (FL--NE, E, S, SW1, NW2) with thick yellow arrows (better defined in Figure \ref{fig_2panel}). We note that there might be some line--of--sight filaments, which are less extended in projection and have poorly resolved alignments in the IRAM 30m 1.2 mm continuum image. However, we expect higher resolution maps to resolve more of these kinds of filamentary structures (Fig. \ref{fig_2panel}). Contours are 20 mJy\,beam$^{-1}$ $\times$ [1, 5, 20] (1$\sigma$= 7 mJy\,beam$^{-1}$). 
}
\label{fig_rgb}
\end{figure}

\clearpage

\begin{figure}
\hspace{-2.8cm}
\vspace{-1.5cm}
\resizebox{\columnwidth}{!}{\includegraphics{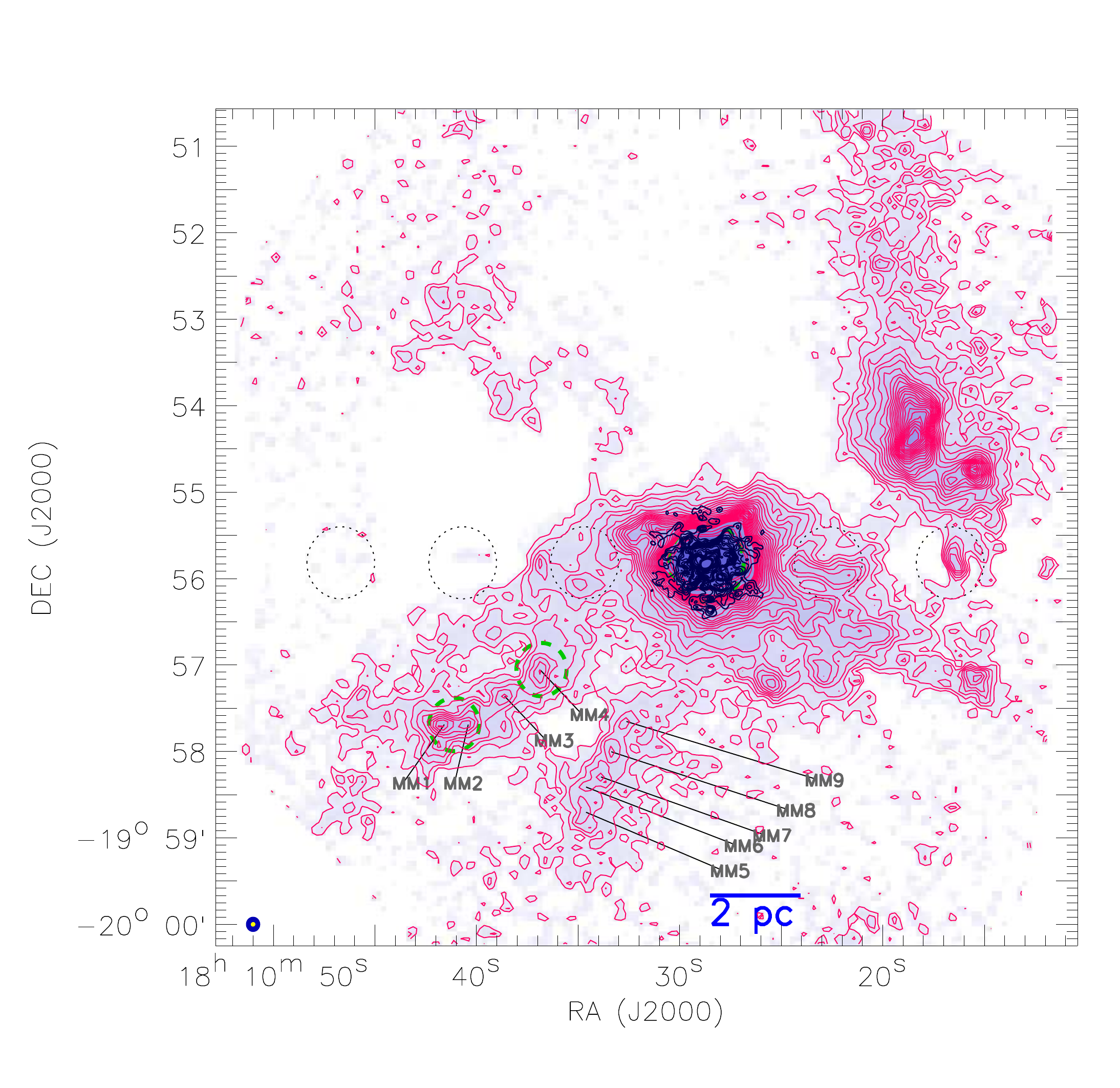}}
\caption{The IRAM 30m 1.2 mm continuum image (magenta contours, blue color scale) and the combined IRAM 30m + SMA image (dark blue contours). The magenta contours are 15 mJy\,beam$^{-1}$ $\times$[1, 2, 3, 4, 5, 6, 7, 8, 9, 10, 11, 12, 13, 14, 15, 16, 17, 18, 19, 20, 21, 22, 23, 24, 30, 36, 42] (1$\sigma$= 7 mJy\,beam$^{-1}$).
The dark blue contours are 16 mJy\,beam$^{-1}$ $\times$ [1, 2, 3, 4, 5, 6, 7, 8, 9, 10, 12, 14, 16, 20, 60] (1$\sigma$= 5 mJy\,beam$^{-1}$). The beam of the IRAM 30m image ($\theta_{maj}$$\times$$\theta_{min}$=10$''$.5$\times$10$''$.5) and the synthesized beam of the combined IRAM 30m+SMA 1.2 mm continuum image ($\theta_{maj}$$\times$$\theta_{min}$=3$''$.4$\times$3$''$.1) are shown by the blue and yellow filled circle, respectively. 
This figure demonstrates the filamentary morphology of the molecular cloud, with embedded massive molecular clumps. 
The identified massive molecular clumps in FL--E and FL--S are labeled by MM 1--9. 
The dashed circle around the dark blue contours represents the primary beam coverage of the SMA 1.3 mm observations; two dashed circles around MM 1, 2 (P1 Region) and MM 4 (P2 Region) represent the primary beam coverages of the SMA 0.87 mm observations. The dotted circles mark the region that are potentially affected by repetitive artifacts (Section \ref{chap_obs}). }
\label{fig_contour}
\end{figure}

\clearpage

\begin{figure}
\hspace{-2.5cm}
\includegraphics[scale=0.45]{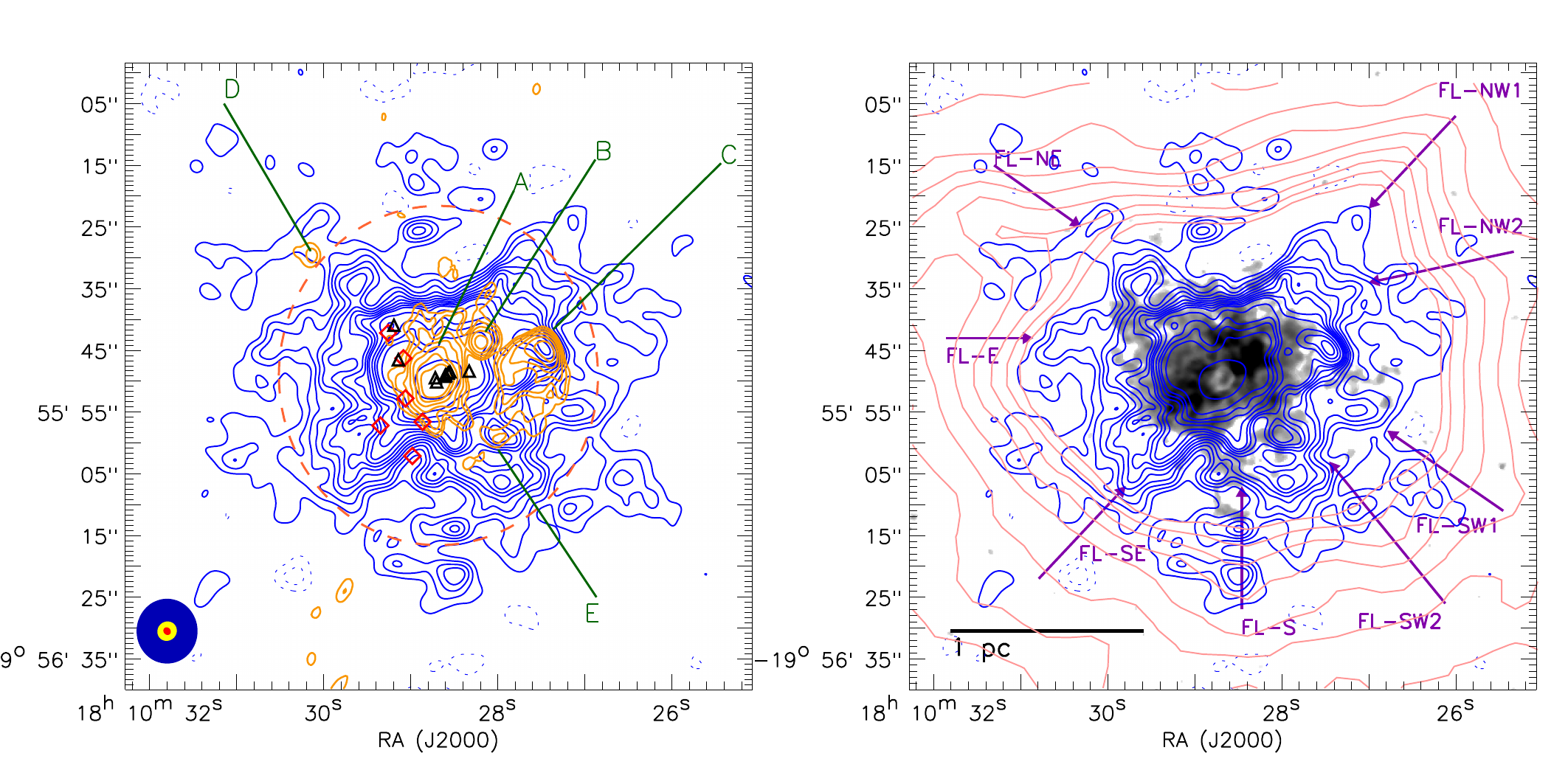}
\caption{The IRAM 30m 1.2 mm continuum image (magenta contours in right panel), the combined IRAM 30m + SMA image (blue contours), the VLA/EVLA 3.6 cm continuum image (orange contours in left panel), and the VLA/EVLA CS (1--0) velocity integrated map (grey scale; see also Liu, Zhang, \& Ho 2011). The (synthesized) beam of the IRAM 30m image, the combined image, and the CS (1--0) image are shown by the blue, yellow, and red filled circle, respectively. A scale bar indicating the 1 pc projected length is shown, with the assumption of a 6 kpc line--of--sight distance. The dashed circle represents the primary beam of the SMA 1.3 mm observations; the UC H\textsc{ii} regions in the SMA primary beam are labeled (A--E) with green arrows (see also Ho, Klein, \& Haschick 1986). The black triangles mark the detections of water masers (Hofner \& Churchwell 1996). The red diamonds mark the 5 previously identified (Liu, Ho, \& Zhang 2010, Section 3.3.1) outflow sources in the intermediate velocity range, and additionally mark the location of the outflow B$_{4}$ and R$_{4}$.  The purple arrows indicate the parsec scale protrusions (FL--NE, E, SE, S, SW1, SW2, NW1, NW2). The blue contours are 16 mJy\,beam$^{-1}$ $\times$ [-2, -1, 1, 2, 3, 4, 5, 6, 7, 8, 9, 10, 12, 14, 16, 20, 60] (1$\sigma$= 5 mJy\,beam$^{-1}$); the orange contours are 3 mJy\,beam$^{-1}$ $\times$ [1, 2, 4, 8, 16, 32, 64] (1$\sigma$= 0.3 mJy\,beam$^{-1}$); the magenta contours are 40 mJy\,beam$^{-1}$ $\times$ [1, 2, 3, 4, 5, 6, 7, 8] (1$\sigma$= 7 mJy\,beam$^{-1}$). The negative contours are shown in dotted lines. The CS distributed in projection against the bright UC H\textsc{ii} regions show negative flux due to strong absorption. We avoid integrating the negative flux while generating the CS map to emphasize the structures in emission. 
}
\label{fig_2panel}
\end{figure}

\begin{figure}
\hspace{-2.5cm}
\includegraphics[scale=0.75]{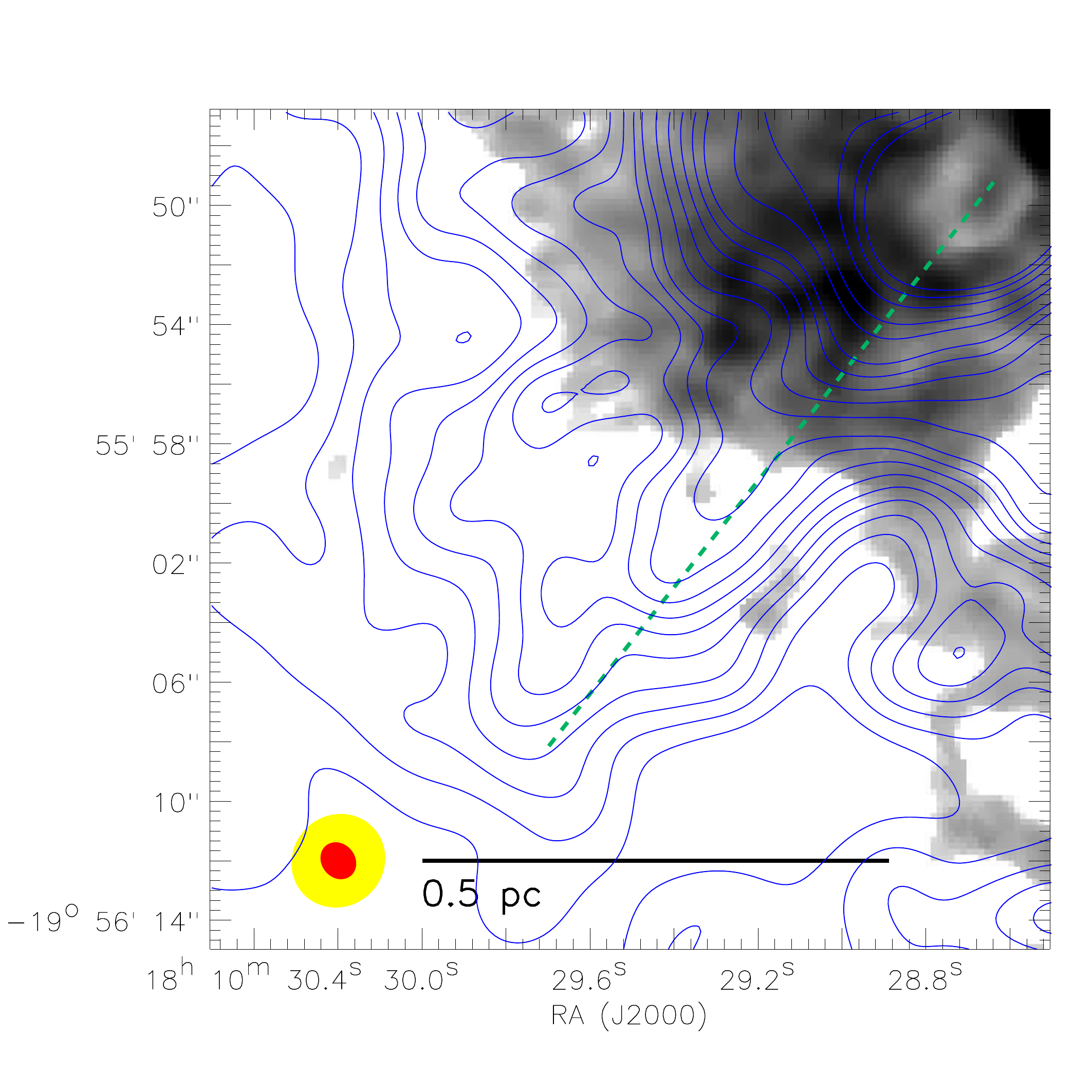}
\caption{The combined IRAM 30m + SMA image (blue contours) and the VLA/EVLA CS (1--0) velocity integrated map (grey scale). The (synthesized) beam of the combined image and the CS (1--0) image are shown by the yellow and red filled circles, respectively. The blue contours are 16 mJy\,beam$^{-1}$ $\times$ [1, 2, 3, 4, 5, 6, 7, 8, 9, 10, 12, 14, 16, 20, 60] (1$\sigma$= 5 mJy\,beam$^{-1}$). The green dashed line show the PV cut as well as a proposed plane of rotation in Liu et al. 2010, Liu, Zhang \& Ho (2011). The dashed line is 25$''$ (0.75 pc) in length, and the top right end of this line corresponds to the zero angular offset. We note a continuation of an elongated CS structure to the 1 mm protrusion along the dashed line. We also note two more elongated CS structures in the north of the previously mentioned one.}
\label{fig_filament}
\end{figure}

\clearpage

\begin{figure}
\hspace{-2.5cm}
\resizebox{\columnwidth}{!}{\includegraphics{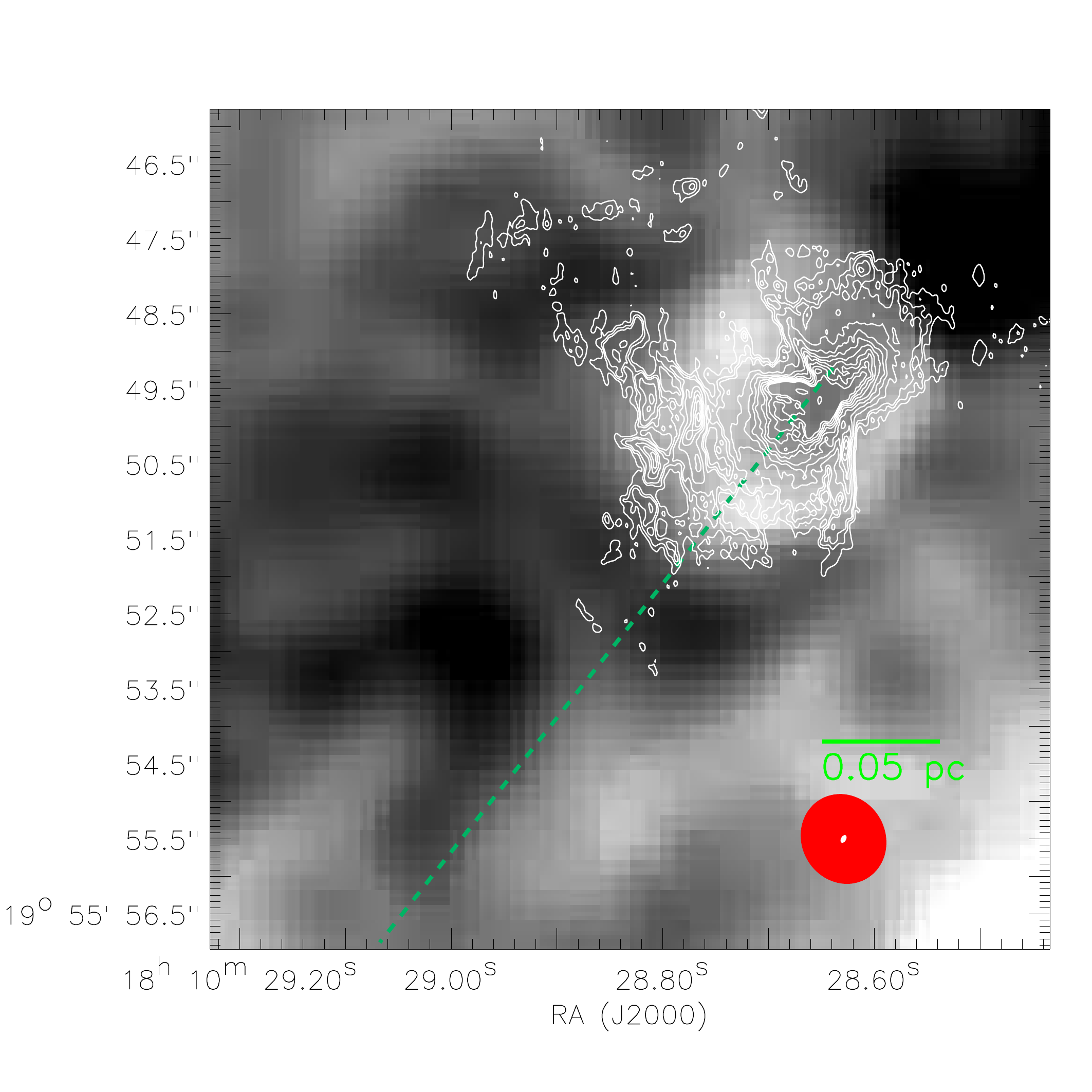}}
\caption{The VLA/EVLA CS (1--0) velocity integrated map (gray scale) and the VLA 1.3 cm continuum map (white contours; see also Sollins et al. 2005 and Sollins \& Ho 2005). The synthesized beam of the CS map and the 1.3 cm continuum map are shown by the red and white filled circles, respectively. Contour levels are  0.4 mJy\,beam$^{-1}$ $\times$ [1, 2, 3, 6, 9, 12, 15, 18, 21, 24, 27, 54] (1$\sigma$= 0.1 mJy\,beam$^{-1}$). The green dashed line show the PV cut as well as a proposed plane of rotation in Liu et al. 2010, Liu, Zhang \& Ho (2011).}
\label{fig_freefree}
\end{figure}

\clearpage

\begin{figure}
\hspace{-2.5cm}
\resizebox{\columnwidth}{!}{\includegraphics{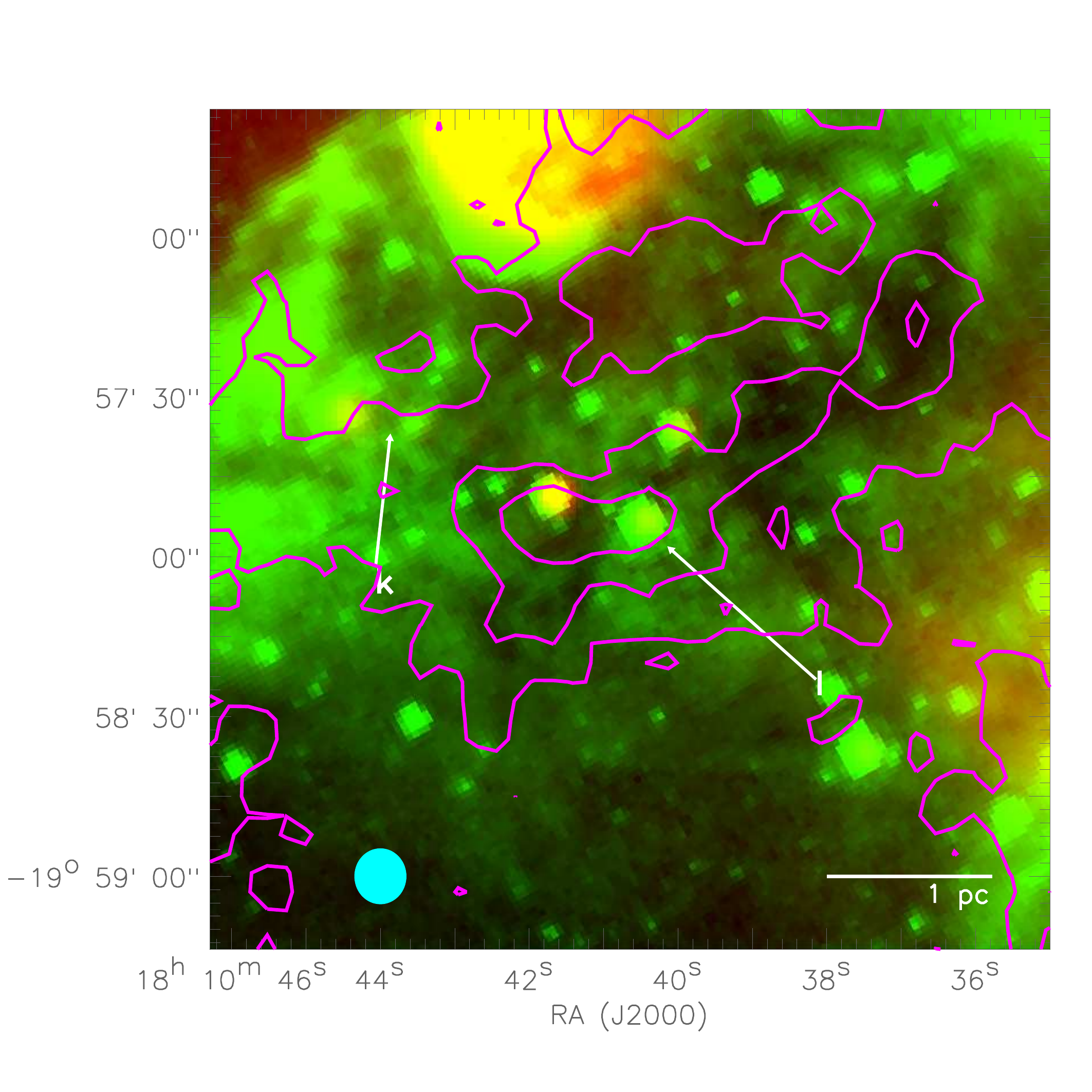}}
\vspace{-0.5cm}
\caption{The IRAM 30m 1.2 mm continuum image (magenta contours), the \textit{Spitzer} MIPSGAL survey 24 $\mu$m image (red), and the \textit{Spitzer} GLIMPSE survey 8 $\mu$m image (green). The magenta contours are [20, 60, 100] mJy\,beam$^{-1}$.
The beam of the IRAM 30m image is shown by the blue filled circles. Some infrared dark regions with complicated structures are consistently traced by the 1.2 mm emission.}
\label{fig_flrg}
\end{figure}

\clearpage

\begin{figure}
\hspace{-2.5cm}
\resizebox{\columnwidth}{!}{\includegraphics{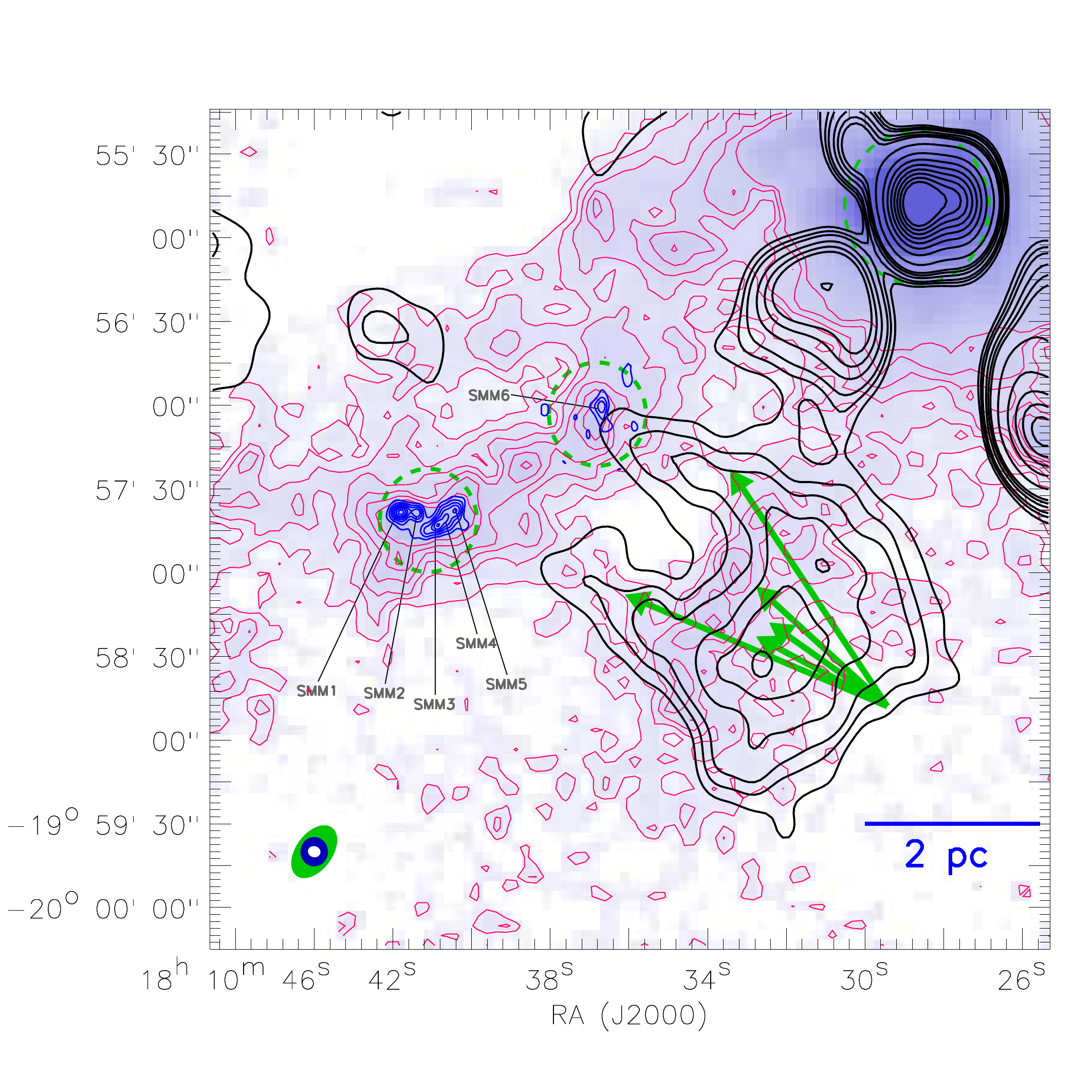}}
\vspace{-1.5cm}
\caption{The IRAM 30m 1.2 mm continuum image (magenta contours, blue color scale), the SMA 0.87 mm continuum image (blue contours; bmaj$\times$bmin = 4$''$.8$\times$3$''$.7, BPA=-5.0$^{o}$), and the VLA 20 cm continuum image (black contours). The magenta contours are 15 mJy\,beam$^{-1}$ $\times$[1, 2, 3, 4, 5, 6] (1$\sigma$= 7 mJy\,beam$^{-1}$). The blue contours are 10 mJy\,beam$^{-1}$$\times$[1, 2, 3, 4, 5, 6, 7] (1$\sigma$= 3.4 mJy\,beam$^{-1}$).
The black contours are 3 mJy\,beam$^{-1}$ $\times$ [2, 3, 4, 6, 8, 10, 20, 30, 40, 50, 60, 70, 80]. The beam of the IRAM 30m image, the synthesized beam of the SMA 0.87 mm continuum image, and the synthesized beam of the 20 cm continuum image are shown by the blue, white, and green filled circles, respectively. The dashed circle in the top right represents the primary beam coverage of the SMA 1 mm observations; two dashed circles around MM 1, 2 (P1 Region)  and MM 4 (P2 Region) represent the primary beam coverages of the SMA 0.87 mm observations.  The suggested illumination pattern of the massive cluster o can be illustrated from the green arrows. We note the H\textsc{ii} region associated with the infrared bright massive cluster j (Figure \ref{fig_rgb}) is detected. The identified submillimeter cores are labeled by SMM 1--6.}
\label{fig_ionized}
\end{figure}

\clearpage

\begin{figure}
\resizebox{\columnwidth}{!}{\includegraphics{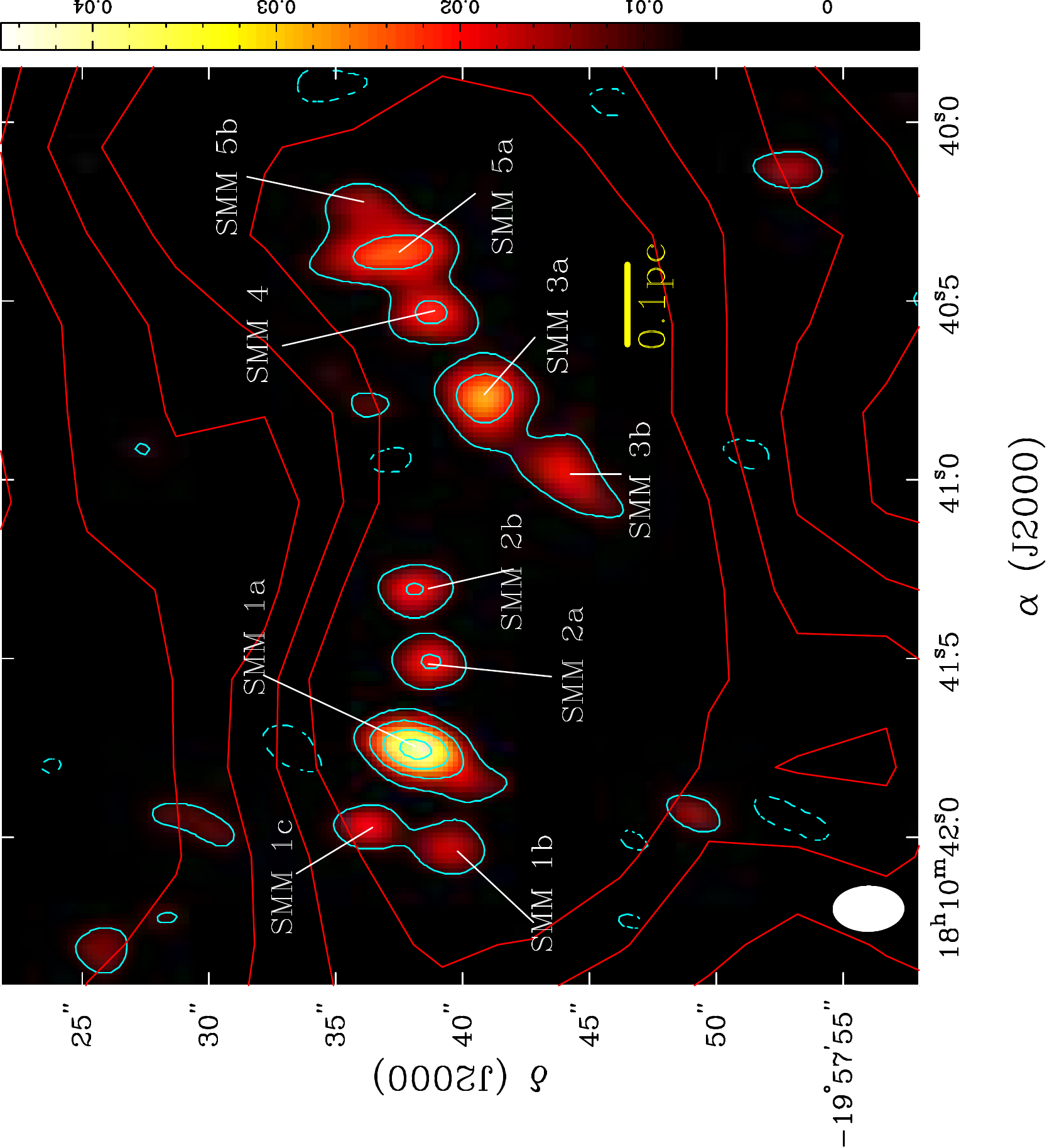}}
\caption{The IRAM 30m 1.2 mm continuum image (magenta contours), and the \textit{high resolution} SMA 0.87 mm continuum image (blue contours and color scale; bmaj$\times$bmin = 2$''$.9$\times$1$''$.9, BPA=0.9$^{o}$) in the P1 region.
The magenta contours are 15 mJy\,beam$^{-1}$ $\times$[1, 2, 3, 4, 5, 6]. The blue contours are 10 mJy\,beam$^{-1}$$\times$[-1, 1, 2, 3, 4] (1$\sigma$= 3.4 mJy\,beam$^{-1}$).
The synthesized beam of the \textit{high resolution} SMA 0.87 mm continuum image is shown in the bottom left corner. The identified submillimeter cores are labeled by SMM 1--5 (a, b, c). These submillimeter cores seem to be aligned linearly.}
\label{fig_condensations}
\end{figure}

\clearpage

\begin{figure}
\hspace{-2.5cm}
\resizebox{\columnwidth}{!}{\includegraphics{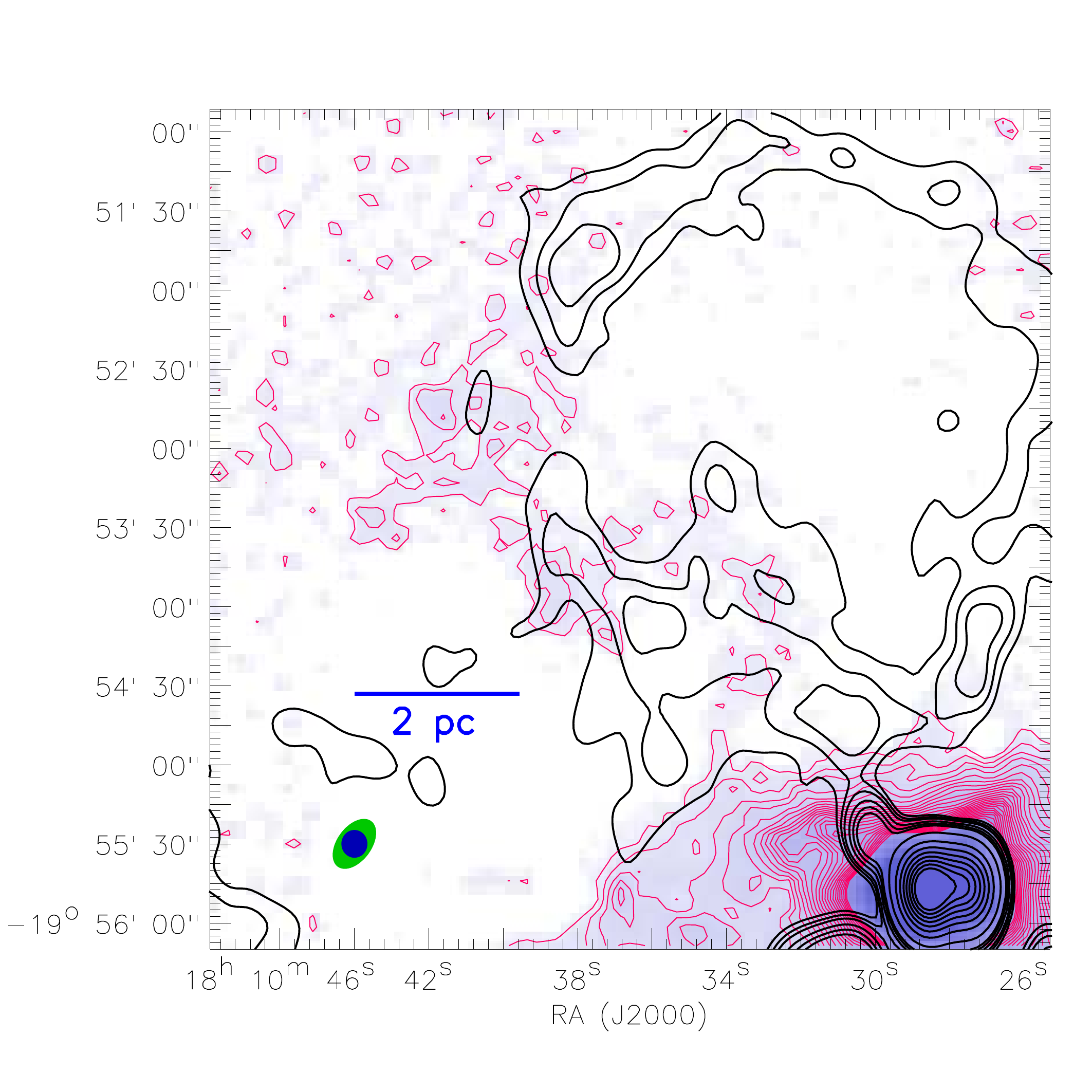}}
\caption{The IRAM 30m 1.2 mm continuum image (magenta contours, blue color scale) and the VLA 20 cm continuum image (black contours). The magenta contours are 15 mJy\,beam$^{-1}$ $\times$[1, 2, 3, 4, 5, 6].
The black contours are 3 mJy\,beam$^{-1}$ $\times$ [2, 3, 4, 6, 8, 10, 20, 30, 40, 50, 60, 70, 80]. The beam of the IRAM 30m image and the synthesized beam of the 20 cm continuum image are shown by the blue and green filled circles, respectively.}
\label{fig_bubble}
\end{figure}


\begin{thebibliography}{}





\bibitem[Beuther et al.(2004)]{2004ApJ...616L..23B} Beuther, H., et al.\ 2004, \apjl, 616, L23



\bibitem[Carroll et al.(2009)]{2009ApJ...695.1376C} Carroll, J.~J., Frank, A., Blackman, E.~G., Cunningham, A.~J., \& Quillen, A.~C.\ 2009, \apj, 695, 1376

\bibitem[Caswell et al.(1975)]{1975A&A....45..239C} Caswell, J.~L., Murray, J.~D., Roger, R.~S., Cole, D.~J., \& Cooke, D.~J.\ 1975, \aap, 45, 239 


\bibitem[Curry(2000)]{2000ApJ...541..831C} Curry, C.~L.\ 2000, \apj, 541, 831 

\bibitem[Downes et al.(1980)]{1980A&AS...40..379D} Downes, D., Wilson, T.~L., Bieging, J., \& Wink, J.\ 1980, \aaps, 40, 379


\bibitem[Fish et al.(2005)]{2005ApJS..160..220F} Fish, V.~L., Reid, M.~J., Argon, A.~L., \& Zheng, X.-W.\ 2005, \apjs, 160, 220





\bibitem[Galv{\'a}n-Madrid et al.(2008)]{2008ApJ...674L..33G} Galv{\'a}n-Madrid, R., Rodr{\'{\i}}guez, L.~F., Ho, P.~T.~P., \& Keto, E.\ 2008, \apjl,674, L33

\bibitem[Galv{\'a}n-Madrid et al.(2009)]{2009ApJ...706.1036G} Galv{\'a}n-Madrid, R., Keto, E., Zhang, Q., Kurtz, S., Rodr{\'{\i}}guez, L.~F., \& Ho, P.~T.~P.\ 2009, \apj, 706, 1036

\bibitem[Galv{\'a}n-Madrid et al.(2010)]{2010ApJ...725...17G}  Galv{\'a}n-Madrid, R., Zhang, Q., Keto, E., Ho, P.~T.~P., Zapata, L.~A.,  Rodr{\'{\i}}guez, L.~F., Pineda, J.~E., \& V{\'a}zquez-Semadeni, E.\ 2010, \apj, 725, 17 

\bibitem[Garay \& Lizano(1999)]{1999PASP..111.1049G} Garay, G., \& Lizano, S.\ 1999, \pasp, 111, 1049

\bibitem[Genzel \& Downes(1977)]{1977A&AS...30..145G} Genzel, R., \& Downes, D.\ 1977, \aaps, 30, 145

\bibitem[Girart et al.(2009)]{2009Sci...324.1408G} Girart, J.~M., Beltr{\'a}n, M.~T., Zhang, Q., Rao, R., \& Estalella, R.\ 2009, Science, 324, 1408

\bibitem[Guilloteau et al.(1988)]{1988A&A...202..189G} Guilloteau, S., Forveille, T., Baudry, A., Despois, D., \& Goss, W.~M.\ 1988, \aap, 202, 189   


\bibitem[Ho \& Haschick(1981)]{1981ApJ...248..622H} Ho, P.~T.~P., \& Haschick, A.~D.\ 1981, \apj, 248, 622

\bibitem[Ho et al.(1983)]{1983ApJ...265..295H} Ho, P.~T.~P., Vogel, S.~N., Wright, M.~C.~H., \& Haschick, A.~D.\ 1983, \apj, 265, 295 


\bibitem[Ho \& Haschick(1986)]{1986ApJ...304..501H} Ho, P.~T.~P., \& Haschick, A.~D.\ 1986, \apj, 304, 501

\bibitem[Ho et al.(1986)]{1986ApJ...305..714H} Ho, P.~T.~P., Klein, R.~I., \& Haschick, A.~D.\ 1986, \apj, 305, 714 

\bibitem[Ho et al.(1994)]{1994ApJ...423..320H} Ho, P.~T.~P., Terebey, S., \& Turner, J.~L.\ 1994, \apj, 423, 320 
       
\bibitem[Ho et al.(2004)]{2004ApJ...616L...1H} Ho, P.~T.~P., Moran, J.~M., \& Lo, K.~Y.\ 2004, \apjl, 616, L1       

\bibitem[Hofner \& Churchwell(1996)]{1996A&AS..120..283H} Hofner, P., \& Churchwell, E.\ 1996, \aaps, 120, 283

\bibitem[Jim{\'e}nez-Serra et al.(2010)]{2010MNRAS.406..187J} Jim{\'e}nez-Serra, I., Caselli, P., Tan, J.~C., Hernandez, A.~K., Fontani, F., Butler, M.~J., \& van Loo, S.\ 2010, \mnras, 406, 187 



\bibitem[Keto et al.(1987)]{1987ApJ...318..712K} Keto, E.~R., Ho, P.~T.~P., \& Haschick, A.~D.\ 1987, \apj, 318, 712 

\bibitem[Keto et al.(1988)]{1988ApJ...324..920K} Keto, E.~R., Ho, P.~T.~P., \& Haschick, A.~D.\ 1988, \apj, 324, 920 
 
\bibitem[Keto(1990)]{1990ApJ...355..190K} Keto, E.~R.\ 1990, \apj, 355, 190 
 

\bibitem[Keto(2002)]{2002ApJ...568..754K} Keto, E.\ 2002, \apj, 568, 754


\bibitem[Keto(2003)]{2003ApJ...599.1196K} Keto, E.\ 2003, \apj, 599, 1196

\bibitem[Keto \& Wood(2006)]{2006ApJ...637..850K} Keto, E., \& Wood, K.\ 2006, \apj, 637, 850



 
 
\bibitem[Klaassen et al.(2009)]{2009ApJ...703.1308K} Klaassen, P.~D., Wilson, C.~D., Keto, E.~R., \& Zhang, Q.\ 2009, \apj, 703, 1308




\bibitem[Larson(1985)]{1985MNRAS.214..379L} Larson, R.~B.\ 1985, \mnras, 214, 379

\bibitem[Ledoux(1951)]{1951AnAp...14..438L} Ledoux, P.\ 1951, Annales d'Astrophysique, 14, 438


\bibitem[Li \& Nakamura(2006)]{2006ApJ...640L.187L} Li, Z.-Y., \& Nakamura, F.\ 2006, \apjl, 640, L187

\bibitem[Lis et al.(1998)]{1998ApJ...509..299L} Lis, D.~C., Serabyn, E., Keene, J., Dowell, C.~D., Benford, D.~J., Phillips, T.~G., Hunter, T.~R., \& Wang, N.\ 1998, \apj, 509, 299 

\bibitem[Liu et al.(2010)]{2010ApJ...722..262L} Liu, H. B., Ho,  P.~T.~P., Zhang, Q., Keto, E., Wu, J., \& Li, H.\ 2010, \apj, 722, 262

\bibitem[Liu et al.(2010)]{2010ApJ...725.2190L} Liu, H.~B., Ho, P.~T.~P., \& Zhang, Q.\ 2010, \apj, 725, 2190

\bibitem[Liu et al.(2011)]{2011ApJ...729..100L} Liu, H.~B., Zhang, Q., \& Ho, P.~T.~P.\ 2011, \apj, 729, 100


\bibitem[McKee \& Ostriker(2007)]{2007ARA&A..45..565M} McKee, C.~F., \& Ostriker, E.~C.\ 2007, \araa, 45, 565

\bibitem[Miyama et al.(1987)]{1987PThPh..78.1051M} Miyama, S.~M., Narita, S., \& Hayashi, C.\ 1987, Progress of Theoretical Physics, 78, 1051

\bibitem[Myers(2009)]{2009ApJ...700.1609M} Myers, P.~C.\ 2009, \apj, 700, 1609 

\bibitem[Nakamura \& Li(2007)]{2007ApJ...662..395N} Nakamura, F., \& Li, Z.-Y.\ 2007, \apj, 662, 395


\bibitem[Omodaka et al.(1992)]{1992PASJ...44..447O} Omodaka, T., Kobayashi, H., Kitamura, Y., Nakano, M., \& Ishiguro, M.\ 1992, \pasj, 44, 447    

\bibitem[Ostriker(1964)]{1964ApJ...140.1056O} Ostriker, J.\ 1964, \apj, 140, 1056






\bibitem[Sollins et al.(2005)]{2005ApJ...624L..49S} Sollins, P.~K., Zhang, Q., Keto, E., \& Ho, P.~T.~P.\ 2005, \apjl, 624, L49           

\bibitem[Sollins \& Ho(2005)]{2005ApJ...630..987S} Sollins, P.~K., \& Ho, P.~T.~P.\ 2005, \apj, 630, 987   



\bibitem[Tang et al.(2009)]{2009ApJ...695.1399T} Tang, Y.-W., Ho, P.~T.~P., Girart, J.~M., Rao, R., Koch, P., \& Lai, S.-P.\ 2009a, \apj, 695, 1399

\bibitem[Tang et al.(2009)]{2009ApJ...700..251T} Tang, Y.-W., Ho, P.~T.~P., Koch, P.~M., Girart, J.~M., Lai, S.-P., \& Rao, R.\ 2009b, \apj, 700, 251

\bibitem[Vishniac(1994)]{1994ApJ...428..186V} Vishniac, E.~T.\ 1994, \apj, 428, 186

\bibitem[Wang et al.(2008)]{2008ApJ...672L..33W} Wang, Y., Zhang, Q., Pillai, T., Wyrowski, F., \& Wu, Y.\ 2008, \apjl, 672, L33

\bibitem[Wang et al.(2011)]{2011ApJ...735...64W} Wang, K., Zhang, Q., Wu, Y., \& Zhang, H.\ 2011, \apj, 735, 64

\bibitem[Wang et al.(2010)]{2010ApJ...709...27W} Wang, P., Li, Z.-Y., Abel, T., \& Nakamura, F.\ 2010, \apj, 709, 27

\bibitem[Whitworth et al.(1994)]{1994A&A...290..421W} Whitworth, A.~P., Bhattal, A.~S., Chapman, S.~J., Disney, M.~J., \& Turner, J.~A.\ 1994, \aap, 290, 421

\bibitem[Wiseman \& Ho(1996)]{1996Natur.382..139W} Wiseman, J.~J., \& Ho, P.~T.~P.\ 1996, \nat, 382, 139

\bibitem[Wiseman \& Ho(1998)]{1998ApJ...502..676W} Wiseman, J.~J., \& Ho, P.~T.~P.\ 1998, \apj, 502, 676


\bibitem[Yorke \& Sonnhalter(2002)]{2002ApJ...569..846Y} Yorke, H.~W., \& Sonnhalter, C.\ 2002, \apj, 569, 846





\bibitem[Zhang et al.(2009)]{2009ApJ...696..268Z} Zhang, Q., Wang, Y.,  Pillai, T., \& Rathborne, J.\ 2009, \apj, 696, 268

\bibitem[Zinnecker \& Yorke(2007)]{2007ARA&A..45..481Z} Zinnecker, H., \& Yorke, H.~W.\ 2007, \araa, 45, 481

\end{thebibliography}
\end{document}